\documentclass[a4paper,fleqn,usenatbib]{mnras}

\usepackage{amsmath}    
\usepackage{amssymb}    
\usepackage{graphicx}   

\usepackage{mathptmx}
\usepackage{txfonts}

\usepackage[T1]{fontenc}
\usepackage{ae,aecompl}

\title[Exocomets in Proxima Centauri system]{%
  Exocomets in the Proxima Centauri system and their importance for water transport%
}

\author[R. Schwarz]{%
  R. Schwarz,$^{1}$\thanks{E-mail: richard.schwarz@univie.ac.at}
  {\'A}. Bazs{\'o},$^{1}$
  N. Georgakarakos,$^{2}$
  B. Loibnegger,$^{1}$
  T.~I. Maindl,$^{1}$
	\newauthor D. Bancelin,$^{1}$
  E. Pilat-Lohinger,$^{1}$
  K.~G. Kislyakova,$^{1,3}$
  R. Dvorak,$^{1}$ 
  \newauthor and I. Dobbs-Dixon $^{2}$
\\
$^{1}$Department of Astrophysics, University of Vienna, T{\"u}rkenschanzstrasse 17, A-1180 Vienna, Austria\\
$^{2}$New York University Abu Dhabi, PO Box 129188, Abu Dhabi, UAE\\
$^{3}$Space Research Institute, Austrian Academy of Sciences, Schmiedlstrasse 6, A-8042 Graz, Austria\\
}

\date{Accepted XXX. Received YYY; in original form ZZZ}

\pubyear{2017}

\begin{document}
\label{firstpage}
\pagerange{\pageref{firstpage}--\pageref{lastpage}}
\maketitle

\begin{abstract}

The scenario and efficiency of water transport by icy asteroids and comets are still amongst the most important unresolved questions of planetary systems. A better understanding of cometary dynamics in extrasolar systems shall provide information about cometary reservoirs and give an insight into water transport especially to planets in the habitable zone. The detection of Proxima Centauri-b (PCb), which moves in the habitable zone of this system, triggered a debate whether or not this planet can be habitable. 
In this work, we focus on the stability of an additional planet in the system and on water transport by minor bodies. We perform numerous N-body simulations with PCb and an outer Oort-cloud like reservoir of comets.
We investigate close encounters and collisions with the planet, which are important for the transport of water.
Observers found hints for a second planet with a period longer than 60~days. Our dynamical studies show that two planets in this system are stable even for a more massive second planet ($\sim 12$ Earth masses). Furthermore, we perform simulations including exocomets, a second planet, and the influence of the binary Alpha Centauri.
The studies on the dynamics of exocomets reveal that the outer limit for water transport is around 200~au. In addition we show that water transport would be possible from a close-in planetesimal cloud (1--4~au). From our simulations, based on typical M-star protoplanetary disks, we estimate the water mass delivered to the planets up to 51 Earth oceans.

\end{abstract}




\section{Introduction}

Comets in the Solar System are known to be a natural by-product of planet formation~\citep{Wahlberg14, 
Willacy15}. It is assumed that comets formed beyond the snow-line~\citep{mulders15}, and they have a higher water content than other minor bodies.
Observations of Solar System comets \citep[e.g.,][]{hartogh11, altwegg15} indicate that their deuterium-to-hydrogen (D/H) ratio exceeds the D/H ratio in Earth's oceans.
These results of comet spectroscopy give a hint on water transport in the Solar System and could be relevant for studies of extra solar systems.

In the work of \cite{delafuente17} they studied the object A/2017 U1 and identified its interstellar origin by its positive barycentric energy.
\cite{mamajek17} investigated the kinematics of this ``exocomet'' and showed that it does not appear to be associated with any local exo-Oort clouds (e.g. the triple system $\alpha$ and Proxima Centauri). 

\citet{Beust90} observed a large number of metallic absorption lines in the spectra of $\beta$~Pictoris, which varied on short time scales. This was interpreted as clouds of gas and dust obscuring the light of the star produced by so called ``falling evaporating bodies''.
Such observations gave indications of exocometary activity even before the first exoplanet was confirmed, which leads to the assumption that many planetary systems could host exocomets.

Between 2003 and 2011 spectra of $\beta$~Pictoris were gathered and then analysed by \cite{Kiefer2014} who discovered about 6000 variable absorption features (Ca II H,K) possibly arising from clouds transiting the disk of the star.
The idea of exocomets producing these clouds seems to be the most reasonable for explaining the observations so far.
The same absorption features have been found in observations of many other star systems, such as HD~21620, HD~42111, HD~110411, and  HD~145964, by \citet{Welsh13} who examined absorption profiles from 21 mainly young ($\sim 5$ Myr) A-type stars with circumstellar disks.
One can assume that these systems are still in the phase of planet formation and the changing gravitational field throws potentially large numbers of icy bodies from the outskirts inwards towards the central star.
The radial velocities measured for the absorption features were in the range of $\pm 100$~km~s\textsuperscript{$-1$}.
From the existence of comets one can assume that there exist also cometary reservoirs in other planetary systems in analogy to the Kuiper-belt or the Oort Cloud.

\cite{Nilsson2010} described the observation of 22 exo-Kuiper-belt candidates at submillimetre wavelength with the LABOCA bolometer. These observations are one example amongst many others for the existence of dusty debris disks around a large fraction of solar type main-sequence stars, inferred from excess far-IR and submillimetre emission, that suggest that leftover planetesimal belts analogous to the asteroid- and comet reservoirs of the solar system are common \citep{Krivov2010,Wyatt2008}. 
 
	Since dust has a limited lifetime, observations which prove its existence lead to the assumption that there are larger asteroidal and/or cometary bodies that continuously renew the amount of dust and form dusty debris disks through collisions. \cite{Wyatt2008} gives a good overview of the evolution and observation of debris disks around main sequence stars and mentions a study by \cite{Jura1998} who state that the presence of warm dust could potentially be a sign for cometary sublimation, which may even dominate the dust population in some of the few hot debris disks.

Assuming that in other planetary systems comets play the same role as in our Solar System, we investigated the possibility of water transport by comets in systems similar to the well discussed system of Proxima Centauri \citep[e.g.,][]{Anglada2016, Coleman2017, Mesa2017}.
At a distance of 1.295 parsecs, the red dwarf Proxima Centauri (M5.5V) is the Sun's closest stellar neighbour and one of the best-studied low-mass stars \citep{seg03}. It is part of the triple star system including the binary Alpha Centauri as shown in the work of \citet{kervella17}. Proxima Centauri is an M-star with a mass of $M = 0.12 \pm 0.015 M_{\mathrm{\odot}}$ separated  on average by 8700~au from the binary system Alpha Centauri A \& B.
\citet{Anglada2016} have announced the discovery of Proxima Centauri b (PCb), a terrestrial planet with a minimum mass of 1.27~$M_{\mathrm{\oplus}}$ orbiting the host star.
The properties of the planet are given in Tab.~\ref{tab1}. PCb is of particular interest because the planet orbits within the habitable zone (HZ) as defined by \cite{kopparapu14}, see Tab.~\ref{tab1}.

\begin{table}
	\centering
	\caption{Orbital parameters of the planet PCb and the location of the HZ in the Proxima Centauri system. The limits of the HZ are calculated using the work of \protect\cite{kopparapu14}. Updates and statistics are presented in the catalogue of exoplanets in binaries and multiple star systems~\citep[][\url{http://www.univie.ac.at/adg/schwarz/multiple.html}]{schwarz16}.}
	\label{tab1}
	\begin{tabular}{lllll}
		\hline \hline
		Name & $a$    & p      & $e$   & $m$ \\
		& (au)   & (days) &       & ($M_{\mathrm{\oplus}}$) \\
		\hline
		PCb  & 0.0485 & 11.186 & 0 -- 0.3 & 1.27 \\
		Habitable zone & 0.045 -- 0.119 & 10.0 -- 43.3 & & \\
		\hline \hline
	\end{tabular}
\end{table}

There still is a great debate whether or not the planet is habitable; e.g., \citet{ribas16} offer a broad overview on that topic. The stellar activity and the fact that PCb is very close to the star might lead to a runaway greenhouse effect and water vapour escape.
Consideration of tidal heating requires an important refinement in the definition of the HZ, as in a general study of \cite{barnes09}.
Additionally for close in planets, like in Trappist-1, the induction heating can melt the upper mantle and enormously increase volcanic activity, sometimes producing a magma ocean below the planetary surface 
\citep{kisly17}. This heating could lead to the complete loss of water from the planet, which may also happen for Proxima Centauri b.

The main goal of this work is to show whether water transport by comets is possible and to assess the contribution of cometary reservoirs with the help of dynamical investigations.
In our study we assume that Proxima Centauri does have comet reservoirs with a considerable water content~\citep{ciesla15}. An observation update by \cite{Anglada17} -- who describe the Atacama Large Millimeter/submillimeter Array (ALMA) detection -- showed a belt of dust orbiting around Proxima Centauri at distances ranging between 1 and 4 au including dust and bodies up to 50~km in size, and signs of an additional outer extremely cold ( $\sim$ 10 K) belt at approximately 30~au which still needs to be confirmed. However, in a recent study \citet{macg18} raise the question whether stellar flares could mimic the dust emission.

The (exo)cometary reservoirs depend on the formation and evolution of the protoplanetary disk.
\citet{vicente05} presented results on the size distribution of circumstellar disks in the Trapezium cluster measured from HST/WFPC2 data.
In their Fig.~8 the disk size is also presented for M stars like Proxima Centauri which would be around 100~au for the spectral type of M5.5V.
The work of \citet{pawellek14} (their Fig.~2) may confirm the result by studying a representative sample of 34 debris disks resolved in various Herschel Space Observatory programs to constrain the disk radii and the size distribution of the dust. These results are important for our initial conditions which we describe in section~\ref{setup2}.

Our primary focus is to study water transport by comets and the related dynamical habitability of PCb. Apart from that, another important dynamical question is: ``Could there be a second planet in the Proxima Centauri system?''
\citet{Anglada2016} announced that there may be a second planet with an orbital period of up to 500 days, but its nature is still unclear due to stellar activity and inadequate sampling.
We address this important issue in section~\ref{res1}, where we analyse the stability of a second planet.

The paper is organized as follows: first we describe the model and methods of our studies (section~\ref{setup}), then we present the results on the stability of an additional planet (section~\ref{res1}), next we show the results of the study on the dynamics of exocomets (section~\ref{res2}), and estimate the lower ande upper limits of the possible water mass transport (section~\ref{watert}). The summary is given in section~\ref{con}.


\section{Models and methods} \label{setup}

\subsection{Stability of an additional planet} \label{setup1}

First, we investigated the possibility of an additional planet (planet c) being stable.
To do this, we used the full 3-body problem (3BP) where we define the star ($m_1$), the planet PCb ($m_2$), and a possible second planet PCc ($m_3$).
We have regarded all the celestial bodies involved as point masses and integrated the equations of motion for the 
stability maps up to $T_{c} =3\cdot 10^6$ periods of PCb ($=10^5$ years).
For our simulations we used the Lie integrator with adaptive step size control \citep[see][]{hanslmeier84, lichtenegger84, eggl10}.
We computed the maximum eccentricity ($e_{\mathrm{max}}$) during the evolution of an orbit to determine its stability and also the region of stable motion around PCb (planet b).
If any planet's (planet b or planet c) orbit becomes hyperbolic ($e_{\mathrm{max}} > 1$) the system is considered to be unstable.
The global results of chaos indicators (e.g., LCI, FLI, RLI, \ldots{}) are in good agreement with the $e_{\mathrm{max}}$ criterion as shown in \citet{schwarz09}.

The initial conditions for PCb were set like in the paper of \citet{Anglada2016} including the uncertainties: $a_{2} = 0.0485$~au, $e_{2} =$ 0, 0.1, 0.2 and 0.3, and an inclination of $i_{2} = 0$ degree.
The second planet (planet c) was started from $a_{3} = 0.06$~au up to $a_{3} = 0.3$~au (for test calculations up to 0.7~au); in addition we varied the initial eccentricity from $e_{3} = 0$ up to $e_{3} = 0.5$. Both planets were always set into coplanar motion.

The estimates of the exoplanet's mass were conducted in the work of \citet{kane17} for three different (line-of-sight) inclination scenarios: $90^{\circ}$ (edge on), $30^{\circ}$, and $10^{\circ}$ (nearly face on).
These inclinations imply a (minimum) mass for the known planet of 1.27, 2.54, and 7.31 $M_{\mathrm{\oplus}}$, respectively. If PCb is coplanar to the discovered dust belts, then the true mass of PCb would be 
1.8~$M_{\mathrm{\oplus}}$.
We set the masses of the planets (planet b and c) equal ($m_2 = m_3$) to decrease the number of free parameters and investigated the following masses:
\begin{itemize} 
    \item One mass of PCb (1~$M_{\mathrm{Prox}}$) $m_2 = m_3 = 1.27$~$M_{\mathrm{\oplus}}$,
    \item 5~$M_{\mathrm{Prox}}$ corresponding to $m_2 = m_3 = 6.35$~$M_{\mathrm{\oplus}}$, and
    \item in addition we made tests for 10~$M_{\mathrm{Prox}}$ corresponding to $m_2 = m_3 = 12.7$~$M_{\mathrm{\oplus}}$.
\end{itemize}

The last two masses correspond to super-Earth planets which are quite common.
\citet{süli05} studied the stability of terrestrial planets with a more massive ``Earth''.
They found that stability of the Solar system depends on the masses of the planets, and small changes in these parameters may result in a different dynamical evolution of the planetary system.

\subsubsection*{Radial velocity}

\begin{figure}
    \centering
    \includegraphics[width=8.0cm]{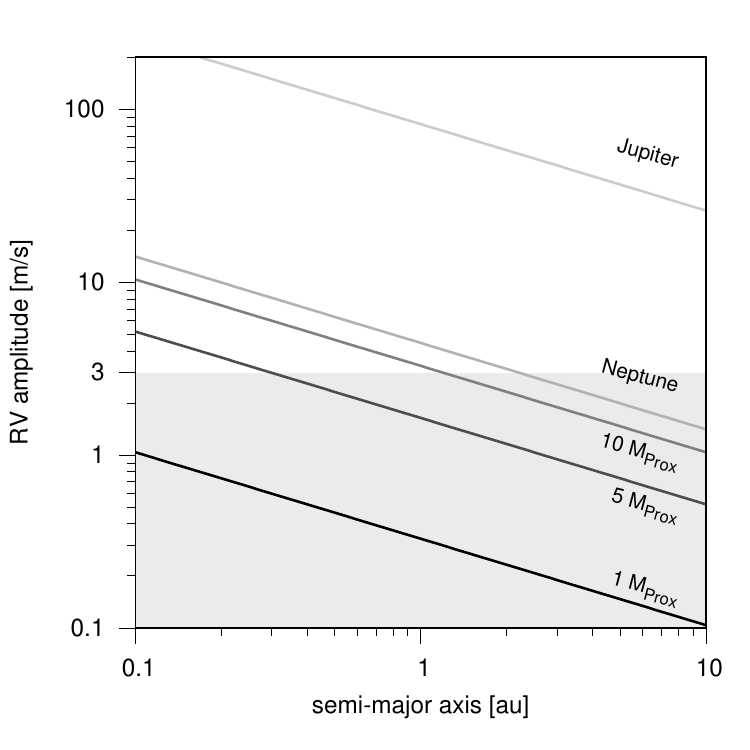}
    \caption{Estimate of the minimum (radial) velocity for Proxima Centauri for the detection of a possible second planet, depending on the mass and the distance of this planet. The lower three masses were used for the stability studies. The grey shaded region represents the observation limit which is RV~$\approx 3$~ms~\textsuperscript{$-1$} \citep{Anglada2016}.}
    \label{fig0b}
\end{figure}

\cite{Anglada2016} wrote, ``that the presence of another super-Earth mass planet cannot yet be ruled out at longer orbital periods and Doppler semi-amplitudes $\lesssim 3$~m~s\textsuperscript{$-1$}, whereas the Doppler semi-amplitude of Proxima b is $\approx 1.4$~m~s\textsuperscript{$-1$}.''
Therefore we made an estimate by the help of Kepler's third law to see which planets would be detectable also in the range of our calculations.
\begin{equation} \label{RV}
    V_{Prox} \approx \left( \frac{m_3}{m_1} \sqrt{ \frac{G m_1}{a_3} } \right).
\end{equation}
Equation~\ref{RV} estimates the radial velocity for Proxima Centauri ($\mathrm{V_{Prox}}$ in m~s\textsuperscript{$-1$}) for a possible second planet, which depends on the mass of the second planet c ($m_\mathrm{{3}}$), the mass of the central star ($m_1$), the gravitational constant and the distance $a_3$ of planet c.
Because of the radial velocity (RV) observations, we do not expect a second planet with a mass larger than a super-Earth.
Fig.~\ref{fig0b} shows the RV amplitude limits in m~s\textsuperscript{$-1$} for the different planet masses by using the limits of~\citet{Anglada2016}. One can see that Neptune-sized planets could only be detected at distances $\le 2$~au.

\subsection{Study on the dynamics of exocomets} \label{setup2}
\subsubsection*{Setup of the dynamical models}
The initial conditions used for the comets in the Proxima Centauri system are based on the setup used in \cite{Loibnegger2017}, who assumed that every system hosting planets also possesses reservoirs of cometary bodies.
These authors investigated the dynamics of comets in the system of HD 10180 and estimated probabilities for collisions
and close encounters with the planets and discussed the possibility of a capture of comets in close-in orbits.
These reservoirs are produced during the phase of planet formation and are later disturbed by e.g. a passing star or the galactic tide which throws the comets towards the central star as studied by \citet{kaib13}.
This was done for the Solar System and the Oort cloud in the work of \citet{fouchard10}.
\citet{fouchard17} introduced a new kind of Oort cloud model; it originates from an Oort cloud precursor in form of a scattered disk with perihelia between Uranus and Neptune.
The objects are launched into orbits with low inclinations and semi-major axes larger than 1000~au.
However, close passing stars may induce comet showers.
\citet{mamajek15} could show that 70000~years ago a low mass binary (HIP~85605) passed the Solar System at around 50000~au (which is at the boundary of the outer Oort cloud).

\begin{figure*}
    \centering
    \includegraphics[width=16.0cm,angle=0]{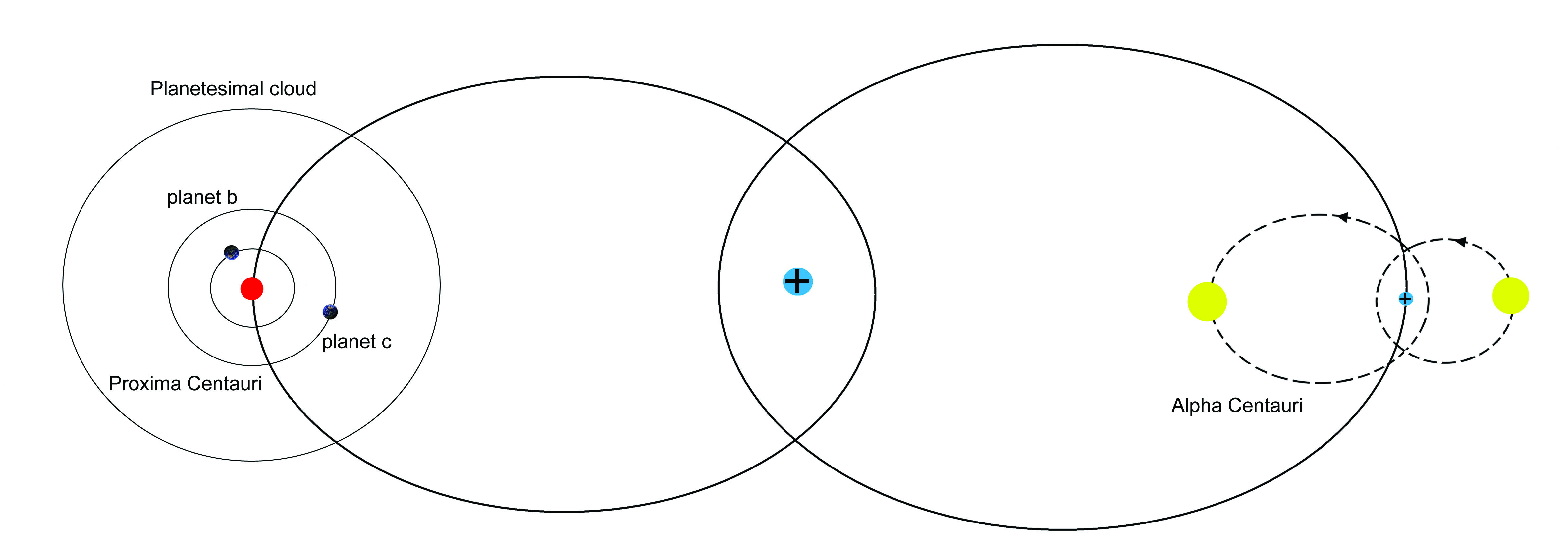}
    \caption{Illustration of the four configurations ({\bf C1}-{\bf C4}) used for our simulations. The mean distance from Alpha Centauri to Proxima Centauri is $a_{\mathrm{bin}} = 8700$~au and the eccentricity $e_{\mathrm{bin}} = 0.5$. The distance of planet b (PCb) is $a_{\mathrm{2}} = 0.0485$~au, while that of planet c (candidate) is $a_{\mathrm{3}} = 0.119$~au at the outer border of the HZ. The six different distances of the planetesimal cloud (only marked as one circle outside the planets' orbits) are listed in this section.}
    \label{fig0}
\end{figure*}

For our investigations we used the following dynamical configurations, which are presented in Fig.~\ref{fig0}:
\begin{description}
    \item [\bf C1:] Star--planet b--planetesimal cloud
    \item [\bf C2:] Star--planet b--planet c--planetesimal cloud
    \item [\bf C3:] Star--planet b--planetesimal cloud--binary
    \item [\bf C4:] Star--planet b--planet c--planetesimal cloud--binary
\end{description}

As mentioned in the introduction, a second planet in the system is possible.
From the work of \citet{kopparapu14} we know that PCb lies at the inner border of the HZ.
The location of planet c is subject to the stability analysis performed in section~\ref{res1}.
Without loss of generality and because of a possible dry PCb, in {\bf C2} and {\bf C4} we set the second planet at 
0.119~au, which is close to the outer border of the habitable zone.

For our simulations of the planetesimal cloud we used 5000 objects. We assumed an Oort-cloud like distribution of inclinations. This choice results in evenly distributed prograde and retrograde comets meaning 
that $0 \le i \le 180^{\circ}$. 
We start the comets on highly elliptic (almost parabolic) orbits. Therefore, we set the initial eccentricity high enough such that the perihelion distance $q < a_2=0.0485$~au in any case; the range of initial eccentricities is $0.95 \le e \le 0.9999$. This is summarized in Fig.~\ref{fig10}. The setup of cometary semi-major axes is also uniform in the intervals defined below:
\begin{description}
	  \item [\bf R0:] 1--4~au
    \item [\bf R1:] 10--20~au
    \item [\bf R2:] 40--50~au
    \item [\bf R3:] 90--100~au 
    \item [\bf R4:] 190--200~au
    \item [\bf R5:] 490--500~au
    \item [\bf R6:] 990--1000~au
\end{description} 

We neglected the dynamical processes that lead to highly eccentric orbits. Such processes could be galactic tides, passing stars, or the gravitational influence of $\alpha$ Centauri.
As mentioned in the introduction, a belt of dust was found around Proxima Centauri at distances ranging between 1 and 4 au \citep{Anglada17}, which is in the region R0. In this paper they found also a hint for a second belt at ~30~au in the ALMA data which is close to the region R2.
Regions {\bf R0} and {\bf R1} represent the short period comets and are the analogue to the Kuiper Belt of our solar system. These short-period comets may have delivered water and other volatiles to Earth and the other terrestrial planets.
Indeed, such exocomets were found in a white dwarf binary system by~\cite{Xu17}.
The outer regions ({\bf R4--R6}) were chosen to investigate the interaction with $\alpha$ Centauri and because the lack of knowledge about the size of the protoplanetary disk.

\subsubsection*{Hill sphere}

Hill's criterion offers a convenient approximation for the stability of planetary systems, like in the work of 
\cite{hamilton92}. The Hill sphere of Proxima Centauri (PC) is defined as the space where its gravitational attraction exceeds that of $\alpha$ Centauri ($\alpha$C).
We use the Hill radius of Proxima Centauri as a valid criterion for the ejection of comets from the star--binary system. The Hill radius for Proxima Centauri is defined as:
\begin{equation}
    R_{\mathrm{Hill}} = a_{\mathrm{bin}} \left( 1 - e_{\mathrm{bin}} \right) \sqrt[3]{\frac{M_{PC}}{3 M_{\alpha C}}} = 1178.3 \: \mathrm{au}.
\end{equation}
The equation depends on the separation $a_{\mathrm{bin}}$ from Proxima to $\alpha$~Centauri, the eccentricity $e_{\mathrm{bin}}$ of Proxima Centauri around $\alpha$~Centauri, and the mass-ratio of Proxima Centauri $M_{PC}$ to the binary $M_{\alpha C}$.
We assume that the outermost region ({\bf R6}) is interior to the Hill radius ($R_{\mathrm{Hill}} \approx 1200$~au).

We studied the restricted N-body problem for an integration time from $T_c = 3.3 \cdot 10^6$ (for the inner regions 
{\bf R0--R3}) up to $T_c = 6.6 \cdot 10^7$ periods of PCb (for the outer regions {\bf R4--R6}).
The equations of motion for all configurations were solved with the Radau method \citep{everhart74, everhart85} from the Mercury integrator package \citep{chambers99}.
Unlike the famous hybrid-symplectic integrator of \citet{chambers99}, the Radau integration method is also well suited for studying binary star systems.
The 15th order Radau method uses Gauss-Radau quadratures to integrate a time series by fitting an empirical curve to force evaluations at several unevenly-spaced points \citep{everhart74}.
Thus, this method is well suited for studying close approaches where a variable step-size adjustment is mandatory.

\subsection{Water transport}

In our simulations we measured the number of:
\begin{enumerate}
    \item Ejected comets
    \item Impacts onto the star
    \item Comets involved in close encounters with the planets b and c
    \item Minimum orbital intersection distance (MOID) of comets with the planets b and c
		\item Direct impacts on the planets b and c
    \item Unique MOID impact events (only the involved comets)
\end{enumerate}

A close encounter with the planet is defined when a comet comes closer than 3 times the Hill radius of PCb.
Comets with semi-major axes $a > R_{Hill} \approx 1200$~au are considered to be ejected.
Possible water transport can be inferred by impacts and by close encounters with the planet.
In case of close encounters the comets can fractionate and collide with the planet as meteor showers like on Earth.
The details of how meteor showers are linked to comets and may have seeded the Earth with ingredients that made life possible is discussed in the work of \citet{jenniskens06}.

We define two impact indicators: direct impacts and minimum orbit intersection distance (MOID).
Direct impacts are retrieved from the simulations for specific impact configurations, that means according to
the real location of the two colliding bodies on their trajectory. This indicator gives a lower estimate of the impact rate, because the measured probabilities are linked to the number of simulated objects.
The second indicator, the MOID algorithm \citep{sitarski68}, is widely used in celestial mechanics to determine whether an object can be a potential danger for the Earth or not.
The MOID corresponds to the closest distance between keplerian orbits of two objects, regardless of their real position on their respective trajectories, considering all possible true anomalies leading to collisions.
It is therefore appropriate to use the MOID for systems like Proxima Centauri where we don't know exactly the parameter space (planets and comets).
For this purpose, we calculated the MOID after the numerical simulations as in \citet{bancelin17} for each comet and estimated the impact probability: each close encounter between the comet and the planet was analyzed by sampling their true anomalies in order to derive impact distances.
Trajectories with $\mathrm{MOID} < R_{\mathrm{Prox}}$ were classified as impacts, where $R_{\mathrm{Prox}}$ corresponds to 1.08 Earth radii and is the estimated radius of PCb for an Earth-like composition \citep{brugger16}.


\section{Results}

\subsection{Investigation of an additional planet} \label{res1}

In this section we present the results of our dynamical investigations concerning a possible second planet in the Proxima Centauri system.
We studied the stability of an additional planet placed from $a_3 = 0.06$~au (orbital period 15 days) up to an initial semi-major axis of $a_3 = 0.3$~au (orbital period 173 days), for eccentricities up to $e_3 = 0.5$.
The second signal mentioned by \citet{Anglada2016} falls into the orbital period range 60--500 days, i.e. in the semi-major axis interval 0.15--0.7~au.

\begin{figure*}
    \centering%
    \includegraphics[width=15.5cm,angle=0]{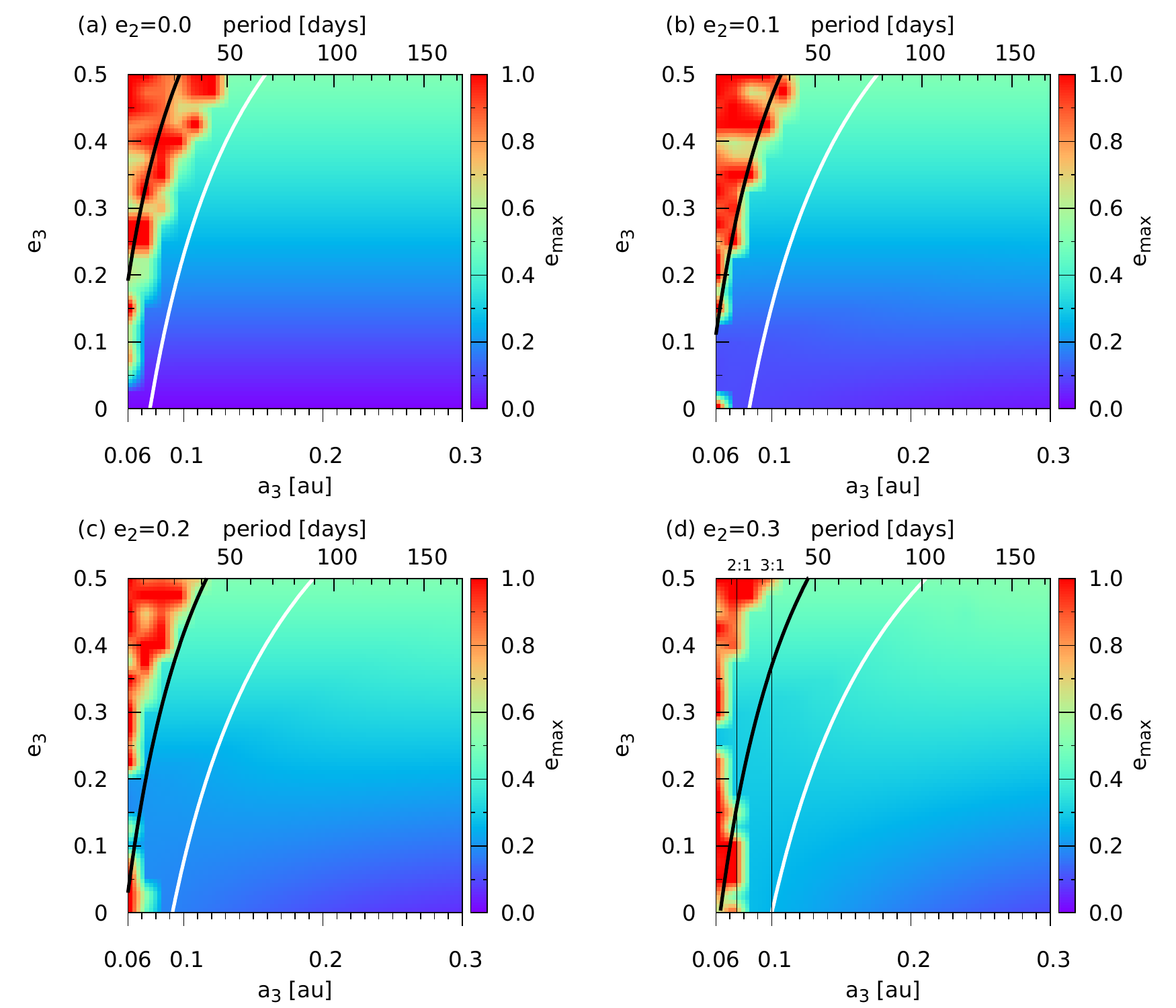}
    \caption{Stability maps of PCc (a possible second planet) presenting the size of the stable region which depends on the initial semi-major axis $a_3$ and the eccentricity of planet c ($e_3$). The masses of the planets are equal $m_2$=$m_3$=5$\cdot M_\mathrm{Prox}$. The 2:1 and 3:1 mean motion resonances are marked in the lower right graph for $e_2$=0.3. The colour code presents the values of $e_\mathrm{max}$. The red regions depict large values of $e_\mathrm{max}$ (corresponding to unstable or chaotic motion), whereas the blue regions represent small ones. Two cases of empirical stability estimates are shown with black and white lines; see text for a description. 
		}
    \label{fig1}
\end{figure*}

Fig.~\ref{fig1} shows the stability maps for the case of $5 M_{\mathrm{Prox}}$.
Each stability map represents different values of eccentricity ($e_2$) starting from $e_2 = 0$ to $e_2 = 0.3$.
In Fig.~\ref{fig1} (lower right graph d) we marked the 2:1 and 3:1 mean motion resonances (MMRs) with PCb.
These resonances are only displayed for the largest eccentricity of PCb $e_2 = 0.3$ (as given in Tab.~\ref{tab1}), but they act also for lower eccentricities.

As a simple indicator for instability we plotted as black curves the condition $Q_{2}=q_{3}$. This means
that above this curve the orbits of the two planets would intersect since the apocenter distance ($Q_2$) for the inner
planet is larger than the pericenter distance ($q_3$) of the outer one. The white curves show the empirical stability estimates from ~\cite{petrovich15}. They developed a new stability criterion by the help of long term integrations ($10^8$ periods) for the three-body-problem discussing exoplanetary systems. 
By applying this criterion we found that when PCc is still located in the HZ (at 0.119 au) it could have eccentricities up to $e_3=0.3$ which confirms our results presented in Fig.~\ref{fig1}.

One can see that the stable region is quite large. To summarize the results we counted the number of stable orbits for each stability map (grid size = $21 \times 21$) which is summarized in Tab.~\ref{tab2}.
We show that the percentage of stable orbits for one $M_{\mathrm{Prox}}$ is very similar for the different eccentricities ($e_2$) and has a mean value of $98.5 \pm 0.5$~\%.
This is also true for $5 M_{\mathrm{Prox}}$ where the mean value is $96 \pm 0.1$~\%.
In addition, we tested the stability for $10 M_{\mathrm{Prox}}$ and did not find any significant difference to the $5 M_{\mathrm{Prox}}$ case.

Finally we can conclude that:
\begin{enumerate}
    \item The largest unstable region was found for the largest eccentricity of PCb ($e_2$=0.3; Fig.~\ref{fig1} lower right graph).
    \item The stability maps with an initial eccentricity of $0 < e_2 < 0.2$ are similar.
    \item The stability maps are varying only slightly with the mass of the planets b and c.
    \item For $0 < e_3 < 0.2$ almost all orbits are stable.
    \item The 2:1 MMR with PCb is visible, but the 3:1 MMR is only visible for higher eccentricities 
		of PCc ($e_3 > 0.4$).
\end{enumerate}

These two MMRs are best visible for a larger eccentricity of the inner planet (PCb), because the width of an MMR depends on the eccentricity of the perturber. When we compare the numerical results with the empirical ones \citep{petrovich15} (white line in Fig.~\ref{fig1}) one can see that all stable orbits are inside the stability criterion. For $e_2=0.2-0.3$ shown in Fig.~\ref{fig1} (lower graphs) we found stable islands in the region above the unstable border (black line). This means there must be an additional mechanism which stabilizes the configuration. 

\begin{figure*}
    \centering%
    \includegraphics[width=15.5cm,angle=0]{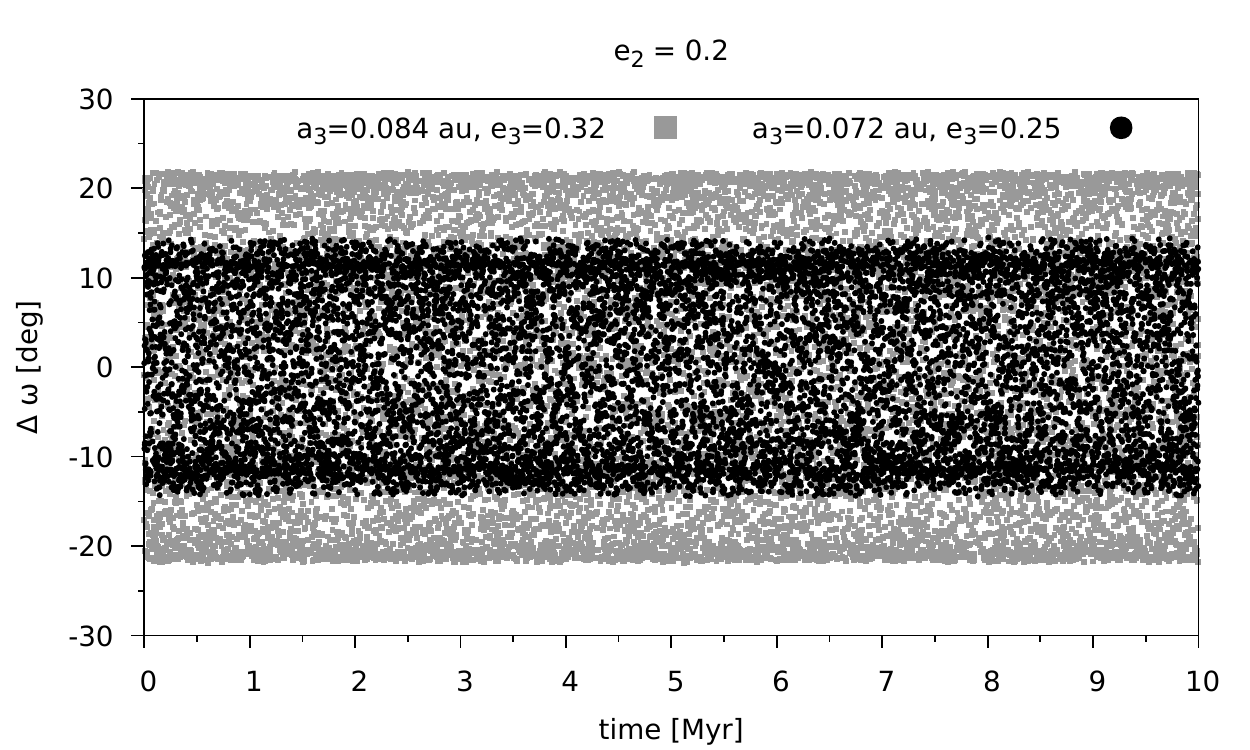}
    \caption{We present an example for the time-evolution of the relative apsidal longitude $\Delta \varpi=\varpi_2-\varpi_3$. The light grey points mark the orbit with the initial orbital elements $a_{3}$=0.084 au, $e_{3}$=0.32 close to the 9:4 MMR; black points depict the orbit with $a_{3}$=0.072 au, $e_{3}$=0.25 close to the 9:5 MMR. It is well visible that for a larger initial eccentricity (grey points) the libration amplitude in $\Delta \varpi$ is larger than for the lower one (black points).}
    \label{fig1a}
\end{figure*}

In addition to resonant arguments for the 2:1 and 3:1 MMR, the relative apsidal longitude 
$\Delta \varpi=\varpi_2-\varpi_3$ is also a critical argument. The libration of $\Delta \varpi$ in an MMR is called apsidal corotation (ACR) as shown in \cite{antoniadou16}.  
If the ACR is stable then neighbouring orbits evolve with $\theta_2$ and $\theta_3$ (resonant angle variables\footnote{$\theta_2 = p\lambda_2 - (p + q)\lambda_3 + q\varpi_2$; $\theta_3 = p\lambda_2 - (p + q)\lambda_3+q\varpi_3$}) showing libration as in Fig.~\ref{fig1a}. Periodic orbits can be approximated by the stationary solutions of an appropriate averaged Hamiltonian which depends on the resonant angle variables presented in a more analytical work of~\cite{beauge03}.
The dynamical exploration of the whole phase space of the resonant problem becomes important, from the point of view of stability of the planetary motion as presented in Fig.~\ref{fig1}. We show in Fig.~\ref{fig1a} an example for the stabilizing ACR mechanism near to the MMRs. This is only valid for small initial 
$\Delta \varpi < 20^{\circ}$.

\begin{table}
    \centering
    \caption{Number of stable orbits and the percentage for all calculated stability maps.}
    \label{tab2}
    \begin{tabular}{cccc} 
        \hline
        $e_2$ & mass                 & number of     & percent of \\
              & ($M_{\mathrm{PCb}}$) & stable orbits & stable orbits \\
        \hline
        0.0 & 1 &432 & 98.0 \\
        0.1 & 1 &432 & 98.0 \\
        0.2 & 1 &435 & 98.6 \\
        0.3 & 1 &439 & 99.5 \\
        \hline
        0.0 & 5 &424 & 96.2 \\
        0.1 & 5 &421 & 95.5 \\
        0.2 & 5 &424 & 96.2 \\
        0.3 & 5 &424 & 96.2 \\
        \hline
    \end{tabular}
\end{table}

\begin{figure}
    \centering
    \includegraphics[width=8.0cm]{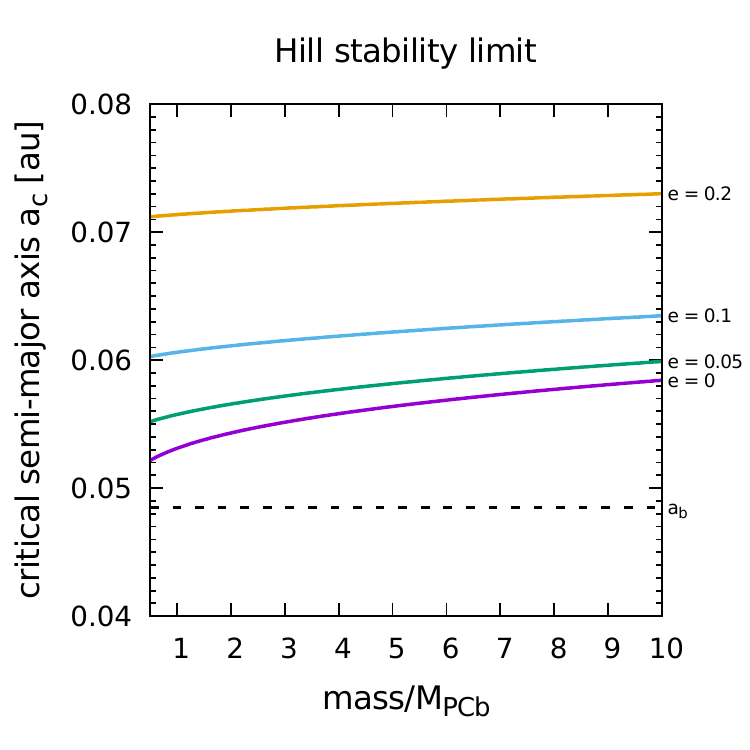}
    \caption{The estimates of the Hill stability limit of two close planets after~\citet{gladman93}.The initial eccentricities of the planets PCb and PCc are always equal ($e_2=e_3$), as shown by the parameter $e$ on the right side of this figure. The x-axis is scaled in units of the mass of PCb.
		 The lowermost dashed line represents the semimajor axis of PCb ($a_b$).}
    \label{fig1b}
\end{figure}

The stability maps in Fig.~\ref{fig1} indicate that Proxima Centauri could indeed host two planets in a compact configuration. To confirm our studies we made use of the Hill stability criterion as done in the work 
of~\citet{gladman93}. This criterion provides Hill stability of two close planets in the planar full three-body problem. In the circular case of the three body problem, the Hill criterion shows a stability limit for PCc for a minimum semi-major axis $a_{c} > 0.053$~au. We represent this criterion depending on the mass and the eccentricity of the planets in Fig.~\ref{fig1b}. The graph displays $a_{c}$ as a function of $M_{PCb}$. The different curves represent the initial eccentricity of the planet PCb, whereas the black line displays the semi-major axis of PCb. A consistent increase of the masses of both planets enlarges the critical separation. 
For larger masses of the planets ($m_2=m_3=5M_{Prox}$) the stability limit shifts to $a_{c} = 0.056$~au, while for eccentric orbits it extends to $a_{c} = 0.062$~au ($e_2=0.1$).
The initial setting of PCc to $a_{c} = 0.119$~au in a circular orbit -- where it is at the outer border of the habitable zone of PC -- for the study on the dynamics of exocomets (next section), is more than two times the required stability limit from PCb.

Finally we have to mention that the stability estimates of \citet{gladman93} and \cite{petrovich15} are a good first hint, but in special configurations and in case of resonances numerical investigations are essential. 

\subsection{Study on the dynamics of exocomets} \label{res2}

\subsubsection{Investigation of one and two planets}

In this study we investigated the influence of the binary $\alpha$ Centauri on the water transport onto PCb and a possible second planet orbiting in the HZ.

In Tab.~\ref{tab3} (upper part) we present the results for {\bf C1} (Star -- planet b -- planetesimal cloud) for the inner three regions of the planetesimal cloud.
The quantity of water transported to PCb decreases, if the number of ejected comets increases and impacts on the stars increases.
The number of ejected comets increases with larger distance to the host star; region {\bf R3} has the largest number.
However, the number of collisions with the star is almost equal.
We have the largest number of close encounters for the inner planetesimal cloud (region {\bf R1}) and less encounters for the outer ones, but the number of involved comets does not differ much (around 80 \%).
In the next step we included a second planet in {\bf C2} which is shown in Tab.~\ref{tab3} (lower part).
It is visible that the region {\bf R1} has more impacts and close encounters than the other regions for both configurations.
When we summarize the number of impacts onto planets b and c we have almost the same number as for {\bf C1} 
({\bf C1:} 7,1,3 and {\bf C2:} 7,3,0).
We can conclude that there is no significant difference in the total number of impacts with the planet between 
{\bf C1} and {\bf C2}.
However, planet b has consistently more impacts in {\bf C2} than planet c, which is valid for all regions as shown in the Tables~\ref{tab3},~\ref{tab4},~\ref{tab5} and is well visible in the impact probability summarized in Tab.~\ref{tab6}.
It seems that the number of impacts will be split if the planets have the same mass without any outer perturber.
The number of MOID impacts is much smaller than the number of close encounters ($\sim$1\% of close encounters lead to an MOID impact), but still many MOID impacts remain; this is valid for both configurations.
We can conclude that the further away the planetesimal clouds from the planets the smaller will be the number of impacts, close encounters and MOID impacts. 

In Tab.~\ref{tab6}, planet b has consistently higher impact probabilities than planet c, because of our initial conditions with highly eccentric comets. 
Investigation of dynamical processes that could lead to such highly eccentric orbits of the comets are beyond the scope of the present study. As we know from the Solar system, highly eccentric cometary orbits are very common \citep{fouchard17}. 
If we included the dynamical evolution of the eccentricities, we would expect that planet c had more impacts than in our simulations. This is caused by the comets with eccentric orbits that at first have only encounters to planet c, before their eccentricity grows large enough to start encounters with planet b. 
To intersect the planet's orbit a comet needs a minimal eccentricity. This minimal value differs only slightly: $e=0.970$ for planet c and $e=0.988$ for planet b (in case of {\bf R0}).
Therefore, as soon as a comet encounters planet c, small perturbations lead to higher eccentricities and open the possibility for encounters with planet b. In consequence the important processes are those external perturbations ($\alpha$ Cen, passing stars and the galactic tide \citet{rickman08}) that can rise cometary eccentricities above $e=0.9$. 

\begin{table}
    \centering
    \caption{Summary of the results for {\bf C1} (planet b) and {\bf C2} (planets b and c). For each region and configuration we calculated 5000 objects.}
    \label{tab3}
    \begin{tabular}{lccc} 
        \hline
        Regions & {\bf R1}& {\bf R2}& {\bf R3}\\
        \hline
        {\bf C1}					& 				&						&					\\
		Ejected comets & 463 & 890 & 1461 \\
		Collisions with the star & 251 & 288 & 288 \\
		Impacts on the {\bf planet b} & 7 & 1 & 3 \\
		Close encounters [$\cdot 10^{5}$] & 9.2 & 2.2 & 1.0 \\
		Comets involved & 4110 & 4158 & 4068 \\
		MOID impacts & 9352 & 2733 & 1748 \\
		Unique MOID impacts  & 173 & 86 & 76 \\
		\hline
		{\bf C2}					& 				&						&					\\
		Ejected comets & 589 & 1001 & 1566 \\
		Collisions with the star & 389 & 542 & 573 \\
		{\bf planet b}					& 				&						&					\\
		Impacts on the planet & 5 & 3 & 0 \\
		Close encounters [$\cdot 10^{5}$] & 9.8 & 2.1 & 2.0 \\
		Comets involved  &  4205 &  4156  &  3940  \\
		MOID impacts & 8608 & 2294 & 2386 \\
		Unique MOID impacts & 286 & 117 & 89 \\
		{\bf planet c}					& 				&						&					\\
		Impacts on the planet & 2 & 0 & 0 \\
		Close encounters [$\cdot 10^{5}$] & 8.2 & 1.7 & 1.8 \\
		Comets involved &  4172  &  3883  &  3521  \\
		MOID impacts & 1398 & 484 & 643 \\
		Unique MOID impacts & 195 & 63 & 33 \\
		\hline	
	\end{tabular}
\end{table}

\begin{figure*}
    \centering
    \includegraphics[width=15.0cm]{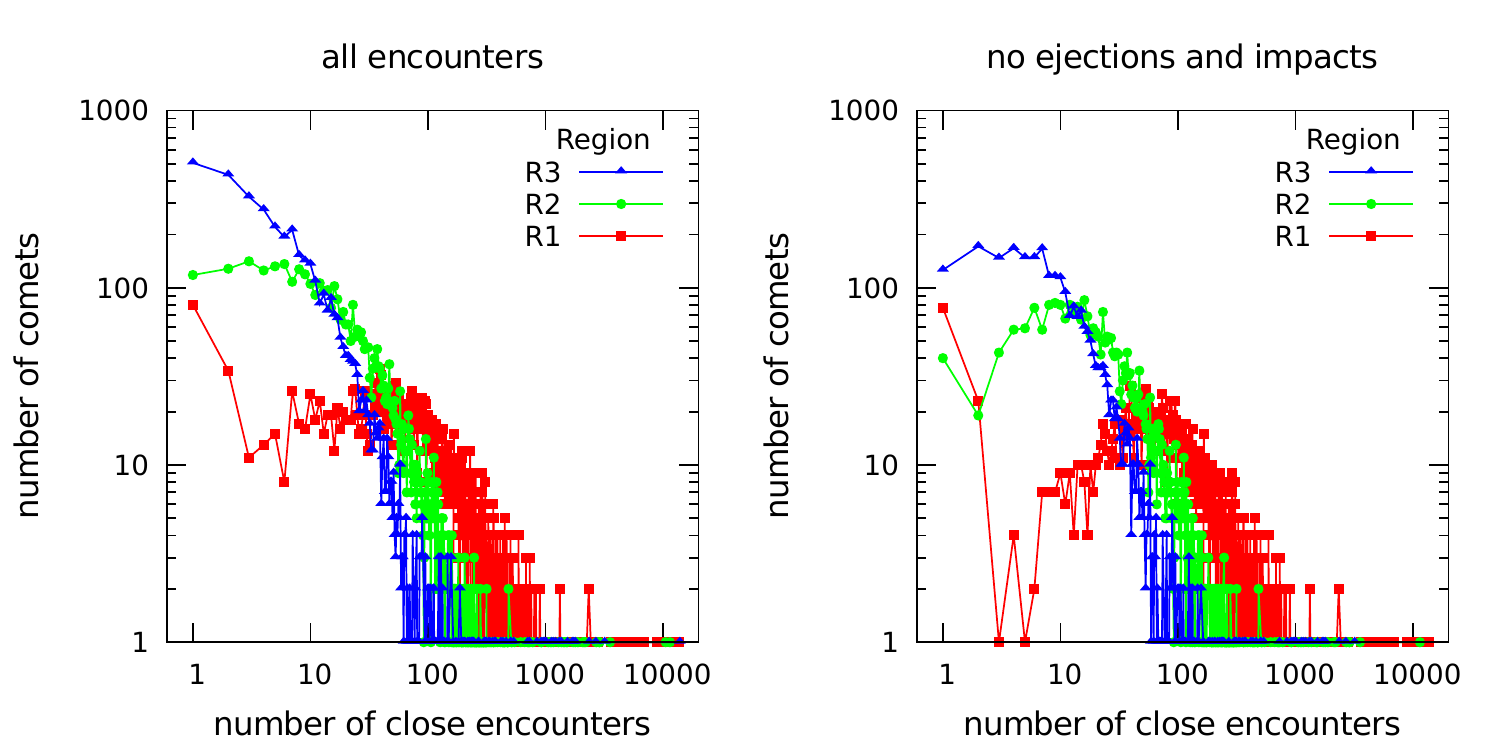}
    \caption{Distribution of the number of close encounters with planet b depending on the number of involved comets for the configuration {\bf C1} with different planetesimal clouds for the inner 3 regions (red line for region {\bf R1:} 10-20~au, green for {\bf R2:} 40-50~au and blue for {\bf R3:} 90-100~au). We counted the number of close encounters for all 5000 objects. It shows that the outer region has more comets with a smaller number of close encounters than the inner ones. The left graph represents all close encounters whereas the right graph shows the same, but without the ejected comets and the ones which have impacts with the host star.}
    \label{fig3}
\end{figure*}

In a next step we prepared a distribution of the number of close encounters by counting the number of involved comets.
In Fig.~\ref{fig3} (left graph) we present the distribution of all close encounters for the configuration {\bf C1} for the inner three regions.
In the right graph of Fig.~\ref{fig3} we present these distributions excluding the ejected comets and the ones which have impacts with the host star. When we compare these two graphs we can conclude that the distribution is almost the same for the inner region ({\bf R1}). There are differences in the distribution for low number of close encounters (1--10 encounters per comet) for all regions.

One can see that many of the involved comets only have relatively few close encounters.
A large number of close encounters implies that the comet would be disrupted and/or de-volatilized and would not contribute to water transport to the planet.

In addition the distribution shows that for region {\bf R1} the number of involved comets is almost constant.
For the regions {\bf R2} and {\bf R3} the number of close encounters rises at the left edge, which means that most comets have only a few close encounters with the planet.

\begin{figure*}
    \centering
    \includegraphics[width=15.0cm]{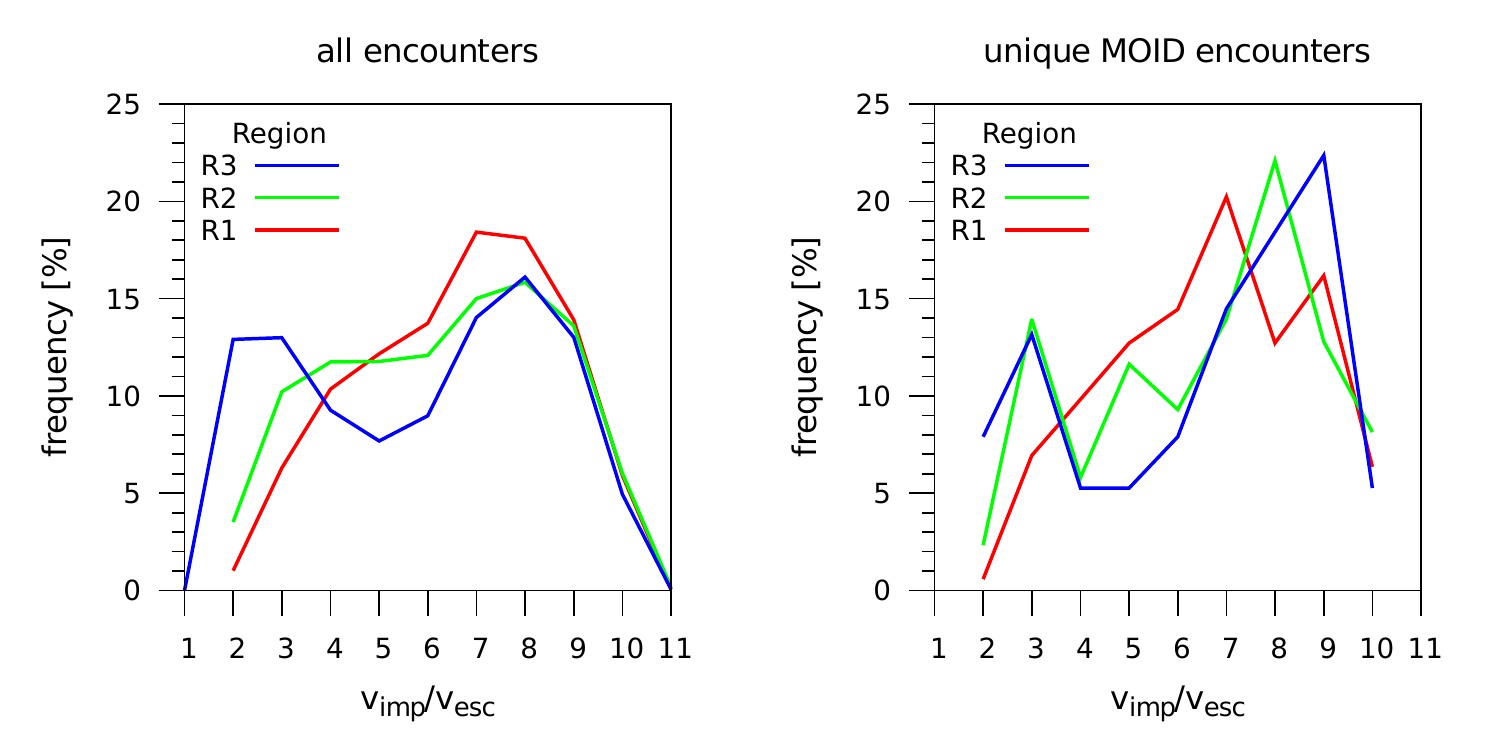}
		\includegraphics[width=14.0cm]{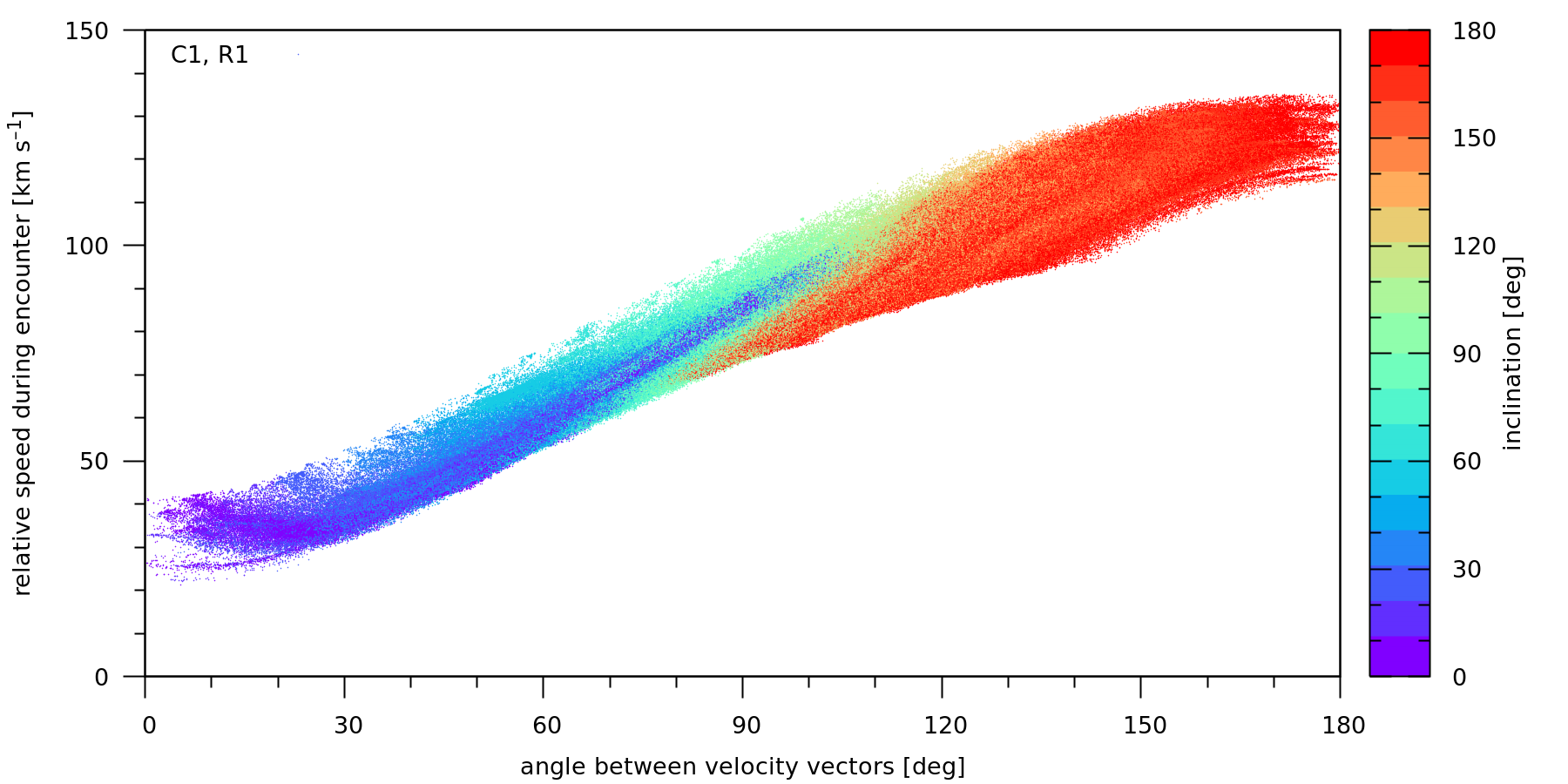}
    \caption{Histogram of the impact velocity for different regions of the planetesimal cloud. Left graph: all close encounters; right graph only taking into account the very first MOID encounter. The lower graph presents the encounter geometry in terms of the angle between the velocity vectors and the relative speed of the encounter. It shows the correlation between the two quantities, and the inclination in a colour code.}
    \label{fig7}
\end{figure*}

For the estimate of the water transport by comets we have to investigate the impact velocity, which
we present as histograms in Fig.~\ref{fig7}. In the left graph we present the impact velocities for all encounters.
The impact speed is normalized to the escape speed from PCb of 12.1~km~s$^{-1}$.
For region {\bf R3} one can see that the histogram has two maxima: the left one represents the impacts of comets with prograde motion whereas the right peak represents the retrograde comets with larger velocities.
The bottom panel of Fig.~\ref{fig7} depicts the correlation of the relative velocity ($v_{comet}-v_{PCb}$) during encounters with the angle between the two vectors. The colour bar indicates the comet's inclination at the instant of the encounters. It is visible that the lowest relative velocities go together with low angles, that means that both objects move in the same direction (prograde comet motion). On the other hand, the highest relative velocities occur when the angles are opposite to each other (retrograde comet motion). We note that the majority of the encounters with PCb happen near the comet's pericenter. 

Accurate estimates of the amount of water deposited on the target (or in its atmosphere) require detailed collision simulations which are out of scope of this study. However, we can still provide an approximate estimate based on previous simulations involving small wet projectiles that impact planet-size targets. In \cite{bancelin17} we studied the impact of rocky Ceres-size objects with a 15 wt-\% water mass fraction on an Earth-sized planet at velocities ranging from 0.5 to 5 mutual escape velocities $v_\mathrm{esc}$. We find that at collision speeds of 0.5, 1, 3, and 5 $v_\mathrm{esc}$ the water lost to space is 0, 11, 37, and 60\,\% of the initial water, respectively. Hence, there is indication that only lower-velocity impacts ($v_\mathrm{imp}\lesssim 5 v_\mathrm{esc}$) contribute to water transport significantly. Due to the more extreme mass-ratio between comets and planets than between large  asteroids and planets, the maximum collision velocity for retaining a significant amount of water will be actually higher. However, moderate impact speeds have another positive effect. \citet{Kral18} discussed for the Trappist-1 system that the atmospheres of the outer planets would gain more volatiles than are lost by impact erosion for moderate impact velocities.

\subsubsection{Calculations including $\alpha$ Centauri}

\begin{table}
    \centering
    \caption{Summary of the results for {\bf C3} and {\bf C4} where the planetary system is perturbed by the binary system $\alpha$ Centauri. For each region and configuration we calculated 5000 objects. *There are no collisions with $\alpha$ Centauri, only with Proxima Centauri.}
    \label{tab4}
    \begin{tabular}{lcccc} 
        \hline
            Regions & {\bf R0}&{\bf R1}&{\bf R2}&{\bf R3}\\
        \hline
		      {\bf C3} &			   & 				&				 &				\\
		Ejected comets &	98		 & 94 		& 104 	 & 90 \\
		Collisions with the star* & 355	 & 377 & 555 & 567 \\
		Impacts on the {\bf planet b}& 70 & 7 & 5 & 2 \\
		Close encounters [$\cdot 10^{5}$]  & 110.0 & 9.0 & 2.6 & 1.6 \\
		Comets involved & 4470 & 4173 & 4533 & 4043 \\
		MOID impacts & 67093 & 7033 & 4073 & 2762 \\
		Unique MOID impacts & 916 & 264 & 257 & 131 \\
		\hline
		{\bf C4}					& 				&						&					\\
		Ejected comets & 110 & 103 & 134 & 105 \\
		Collisions with the star* & 476 & 467 & 674 & 741 \\
		{\bf planet b}		&			& 				&						&					\\
		Impacts on the planet & 60 & 8 & 1 & 0 \\
		Close encounters [$\cdot 10^{5}$] & 111.9 & 12.7 & 2.0 & 1.3 \\
		Comets involved  & 4471 & 4207 & 4445 & 3809 \\
		MOID impacts & 55821 & 9086 & 1771 & 1446 \\
		Unique MOID impacts & 1487 & 405 & 284 & 117 \\
		{\bf planet c}		&			& 				&						&					\\
		Impacts on the planet & 11 & 1 & 0 & 0 \\
		Close encounters [$\cdot 10^{5}$]& 97.1 & 10.9 & 1.7 & 1.2 \\
		Comets involved & 4521 & 4085 & 4174 & 3749 \\
		MOID impacts & 12972 & 2222 & 304  & 320 \\
		Unique MOID impacts & 1021 & 230 & 85 & 43 \\
		\hline	
	\end{tabular}
\end{table}

\begin{figure*}
    \centering
    \includegraphics[width=12.5cm]{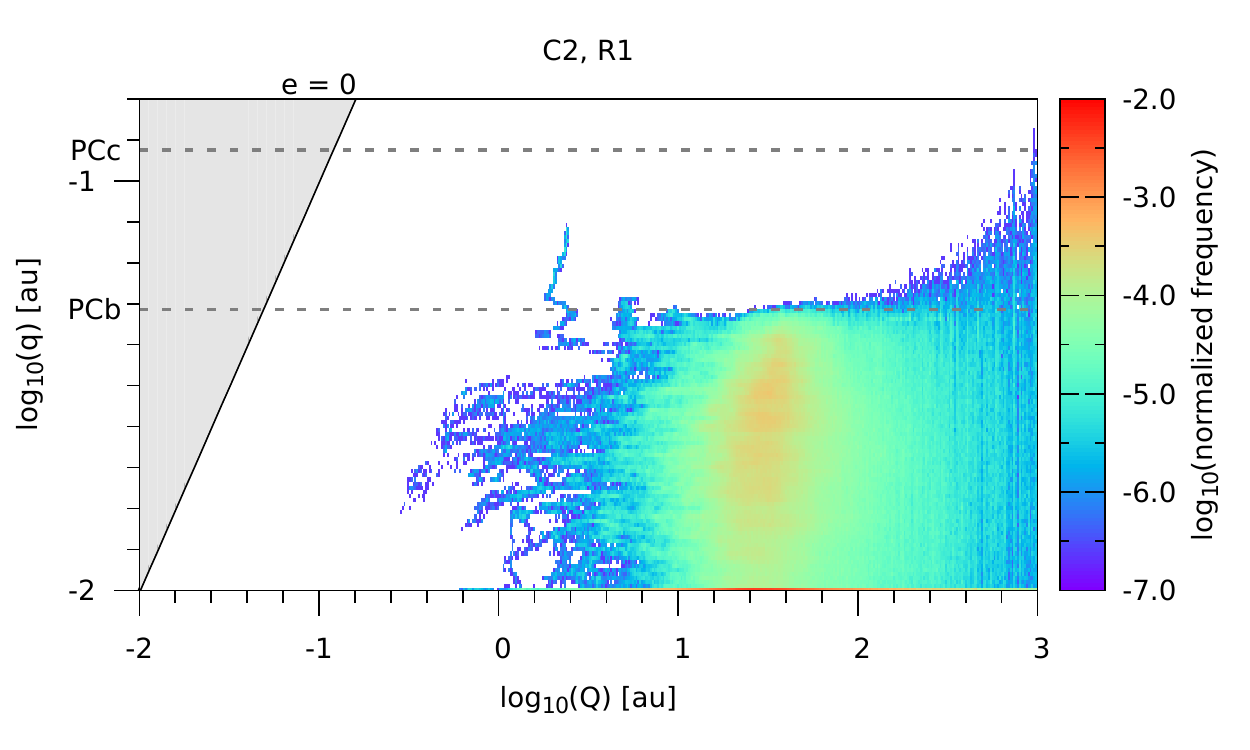}\\
		\includegraphics[width=12.5cm]{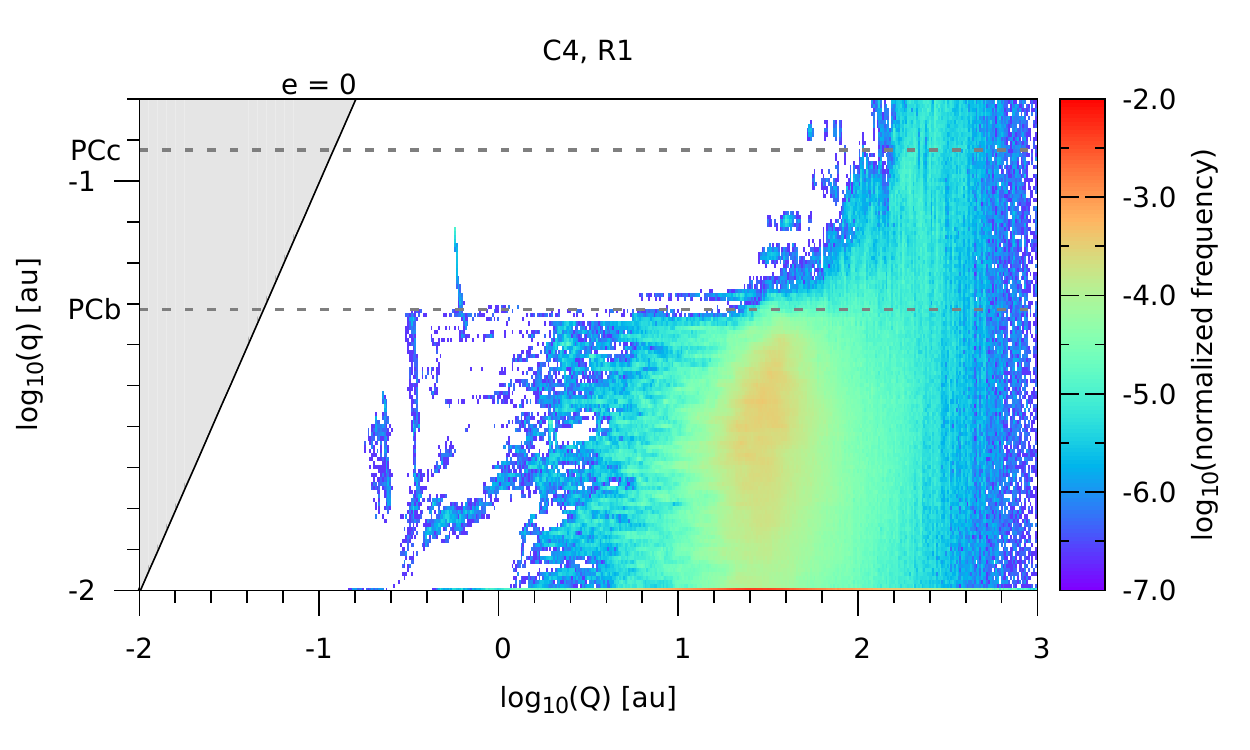}
    \caption{Distribution of the residence times of all involved comets for the configurations {\bf C2} (upper graph) and {\bf C4} (lower graph) for the region {\bf R1}. We plot the logarithms of the aphelion distance $Q$ versus the perihelion distance $q$. The colour code of the figure is the normalized frequency of the comets to reside in a certain cell of the grid. The grey area is inaccessible to comets, because the lower limit for the eccentricity ($e=0$) is a circular orbit with equal perihelion and aphelion distances.}
    \label{fig3a}
\end{figure*} 

We analysed the cometary dynamics by means of residence time maps \citep[after][]{bottke02} in the osculating aphelion distance (Q) -- perihelion distance (q) space (as shown in Fig.~\ref{fig3a}). Each cell gives the normalized frequency of the comets of the map where brighter colours indicate higher residence time. 
Any comet with constant aphelion distance but changing perihelion distance with time (due to close encounters with the planets) appears as a vertical stripe. The opposite behaviour is also possible, i.e. horizontal stripes correspond to a constant perihelion distance but a variable aphelion distance with time. The dashed lines show the perihelion limits for encounters to the planets PCb and PCc. Note that in configuration {\bf C4} many comets are scattered to large distances, beyond the perihelion distance of 
PCc (marked as the upper dashed line at the lower graph in Fig.~\ref{fig3a}). 

This indicates that the binary influences the dynamics of the Proxima Centauri system.
To show the effects of the binary system $\alpha$ Centauri we compared the different configurations ({\bf C1-C4}) which are summarised in Tables~\ref{tab3} and \ref{tab4}. 
First, we can see that the number of ejected comets is smaller for {\bf C3} compared to {\bf C1}, whereas the number of collisions with the star increased for {\bf C3}.
This is also valid when we compare configuration {\bf C4} with {\bf C2}. We have to mention that the ejection distance for {\bf C3} and {\bf C4} is 20000~au (to include $\alpha$ Centauri) instead of approximately 1200~au (Hill radius) before. Due to the restricted simulation time this causes that some hyperbolic comets are far away but not yet considered to be ejected.
The number of impacts with one planet ({\bf C1} and {\bf C3}) is similar, as in comparison of {\bf C2} with {\bf C4} with two planets.
But the number of close encounters for the configurations ({\bf C3, C4}) including $\alpha$ Centauri is slightly higher than without the binary system ({\bf C1, C2}), as well as the number of involved comets.
However, the number of MOID impacts increases significantly when we compare {\bf C3} with {\bf C1} and {\bf C4} with {\bf C2}.
Therefore we can conclude that the influence of $\alpha$ Centauri in ({\bf C3, C4}) on the possible water transport is only visible in the number of unique MOID impact events.

\subsubsection{Investigation of the outer regions}

Tab.~\ref{tab5} shows the dynamical study with a possible planetesimal cloud further out represented by the regions {\bf R4, R5} and {\bf R6}.
These studies were done for {\bf C4} only, which includes $\alpha$ Centauri and a second planet.
For the outer regions we increased the integration time to $T_c$=6.6$\cdot 10^7$ periods of PCb, to make sure to cover several periastron passages of $\alpha$ Centauri.
One can see that the planetesimal cloud for region {\bf R6} (1000 au) is heavily depleted by ejections and collisions with the star.
The number of ejected comets increases with larger apocenter distance ({\bf R4-R6}).
This cometary diffusion happens because comets receive kicks when their pericentres are close to the planet.
These kicks may increase their semi-major axes, and they will be ejected afterwards~\citep{wyatt17}. 
For {\bf R6} we note that there are no close encounters with the planet, because $\alpha$ Centauri is very efficient in scattering comets from the system.
 
Therefore and because of the close encounters, MOID, and impacts we can conclude that from region {\bf R6} no water transport is possible.
The comets from this far away region have only impacts with the star Proxima Centauri.
Region {\bf R5} shows similar results except some close encounters with the planets are possible.

However, region {\bf R4} has less ejected comets and collisions with the star.
We expect some limited water transport because of the number of impacts (three in total), close encounters with the planets and the MOID.
Because of region {\bf R4} we expect that the limit for water transport is around 200~au.
Further intense investigations are necessary to obtain a better estimate of this outer limit.

\begin{table}
    \centering
    \caption{Results for scenario {\bf C4} where the planetary system is perturbed by the binary system $\alpha$ Centauri, but for the outer regions {\bf R4, R5} and {\bf R6}. For each region we calculated 5000 objects. *There are no collisions with $\alpha$ Centauri only with Proxima Centauri.}
    \label{tab5}
	\begin{tabular}{lccc} 
		\hline
		Regions & {\bf R4}&{\bf R5} &{\bf R6}\\
		\hline
		Ejected comets & 493 & 127 & 1368 \\
		Collisions with the star* & 834 & 979 & 1733 \\
		{\bf planet b}					& 		&				&			\\		
		Impacts on the planet & 2 & 0 & 0 \\
		Close encounters 		 	& 8043 & 23 & 0 \\
		Comets involved  			&  155 &  6 & 0 \\
		MOID impacts					&	 169 &  0 & 0 \\
		Unique MOID impacts		&	  25 &  0 & 0 \\
		{\bf planet c}				& 		&			&				\\		
		Impacts on the planet & 1 & 0 & 0 \\
		Close encounters 		 	& 9494 & 19 & 4 \\
		Comets involved  			&  256 & 13 & 1 \\
		MOID impacts					&	  56 &  0 & 0 \\
		Unique MOID impacts		&	  20 &  0 & 0 \\
		\hline	
	\end{tabular}
\end{table}

\subsection{Water transport}
\label{watert}

To estimate the amount of water transported by comets to the planets b and c we concentrate on impacts as the most efficient transport mechanism.
We can estimate how much water may be transported for the different configurations by the following equation:
\begin{equation} \label{eq:watertransport}
    W_{\mathrm{H_2 O}} = M_{\mathrm{tot}} \times wmf \times P(q \le a_P) \times P(I) / M_{\mathrm{H_2 O}}.
\end{equation}
In this expression the different factors are either derived from the simulations -- such as the impact probability $P(I)$ -- or they are model dependent like the water mass fraction ($wmf$).

The first element is the total mass $M_{\mathrm{tot}}$ of each region $R_i$ ($i$ = 1 \ldots 6); these values are given in Tab.~\ref{tab7}. This total mass is calculated by integrating the surface density profile $\Sigma(r) = \Sigma_0 \, (r / 1 \mathrm{AU})^{-\alpha}$ with the parameters ($\alpha, \Sigma_0$).
We assume two different values for the exponent: $\alpha = 1.5$ and $\alpha = 2.2$.
The first value \citep{crida09,ciesla15} comes from the minimum mass solar nebula model for our solar system and results in a radial decrease of the total mass. A newer investigation using Kepler-detected planets of M-dwarfs by \cite{gaidos17} suggests $\alpha = 2.2$, where the surface density profile is much steeper and results in less mass and water for distant regions. The surface density value $\Sigma_0$ = 30~g~cm\textsuperscript{$-2$} describes a mixture of solids plus ice~\citep{hayashi81}, and is valid for regions beyond the system's snow line (at about 1~AU
see \citet{Coleman2017}). Based on these assumptions we expect initial masses for the planetesimal clouds of approx. $10^{-5}-10^{-7} M_{\odot}$. In the paper of \citet{Anglada17} they derived a total dust mass of 0.01 $M_{\oplus}$. 
This low value is caused by dust removal by collisional grinding processes and subsequent removal of the dust by stellar radiation pressure. \citet{macg18} caution that stellar flares from Proxima Centauri could mimic the dust emission. Therefore, it is not clear how reliable the estimates of the dust mass are. As a simplification we assumed that the whole mass of the initial disk had been converted into comets. Over time comets are lost by impacts and ejections, so that in the evolved Proxima Centauri system there is less material available than in the early phase,
and consequently water transport decreases with time.

The water mass fraction $wmf$ was chosen to be around 50 percent following \citet{ciesla15}.
This value is rather high compared to the typical $wmf$ of asteroids (5--10~\%).
Since all our source regions are situated beyond the snow line we can expect this large water content to be reasonable.

In order to determine the amount of the total stored water in each region that is able to be transferred to the planets we need to assess the transport probability $P(q \le a_P)$, i.e. the fraction of comets achieving highly elliptic orbits that take them to pericenter distances $q$ interior to the planet's orbit $a_P$.
From the eccentricity distributions\footnote{Based on data from the JPL Small-Body Database Search Engine, \url{https://ssd.jpl.nasa.gov/sbdb_query.cgi}} of minor bodies in the solar system, we can look for those objects with
$e \geq 0.98$ (this is the lower limit for {\bf R0}), like trans-Neptunian objects (TNOs) 2014~FE72, 2017~MB7, 2005~VX3. 
From these distributions we can infer that this fraction is between 0.01 (TNOs) and 0.3 (LPCs).
However, the latter value is probably biased since only objects from the high-eccentricity tail of the distribution will reach the inner parts of the Solar System and become visible. Therefore we select $P(q \le a_P) = 0.01$ as a more realistic estimate.
Note that comets do not start on such highly eccentric orbits, but there is a dynamical time scale during which
perturbations (by $\alpha$ Centauri, galactic tides, $\ldots$) act. As Proxima Centauri is quite old we neglect this timescale.

The impact probability $P(I)$ is described by two sets of values. We estimate the lower limit of the water mass transport to the planets by using the probabilities for direct impacts. As an upper limit we use the probabilities for the unique MOID impacts. The actual water mass is expected to lie in the range between these two extremal values.

Note that in Eq. (\ref{eq:watertransport}) the transported water mass $W_{\mathrm{H_2 O}}$ is scaled in units of one Earth ocean mass, $M_{\mathrm{H_2 O}} = 1.5 \times 10^{21}$~kg.

In Fig.~\ref{fig3a} we show that comets can approach the host star very closely ($q \le 0.01$~au).
This leads to the question of water (ice) loss by sublimation due to the stellar radiation.
To estimate the sublimation mass flux one can use the method of \citet{jewitt12,bancelin17}. The mass flux averaged over one orbit as well as over the lifetime of a comet (in the simulations) is irrelevant over the simulation time-span.
Sublimation should be maximal during pericenter passage, but at a typical eccentricity $e > 0.99$ the whole passage through the Proxima Centauri habitable zone takes less than about 10 days.
We have to compare this period of time with the orbital periods, which are typically of the order of hundreds of years. Hence these highly eccentric comets spend much time at large distances to the star where the temperature is too low for any significant sublimation. 
\citet{prialnik09} estimated the sublimation time-scales for main-belt comets as larger than $10^8$ years (for temperatures below 100~K). Considering region {\bf R0}, \citet{Anglada17} estimated a dust-temperature of 30~K which would even enlarge the sublimation time-scale. \citet{Haghighipour18} showed that impact craters deeper than 15~m are sufficient to trigger the activity of main-belt comets. This means that such a dust-layer prevents a comet from sublimation effects. Therefore we did not consider this water loss process.

The results for water transport are presented in Tab.~\ref{tab7} (and is summarized in Fig.~\ref{fig9}) for two different exponents $\alpha=1.5$ (minimum mass solar nebula) and $\alpha=2.2$ from newer investigation by \cite{gaidos17}. In case of $\alpha=1.5$ the conservative minimum water mass transport represented by direct impacts can be up to 4 Earth oceans (region {\bf R0}). In contrast to that, the optimistic water transport represented by MOID impacts can rise up to 84 Earth oceans (again region {\bf R0}). For $\alpha=2.2$ we obtained a water mass transport of up to 2.4 and 51 Earth oceans by direct impacts and by MOID impacts, respectively.
We can conclude that the water mass transport is dominated by the regions {\bf R0} and {\bf R1}, while the outer regions contribute a minor fraction.

We remark that our simulations are focused on the water transport to the planet(s). This does not mean that
all the water will stay on the planet. If there is no atmosphere liquid water cannot be preserved. Otherwise, water vapour can be lost from the post impact ejecta as \citet{shu09} has shown for oblique and high velocity impacts. 

\begin{table}
    \centering
    \caption{Summary of the impact probability of direct impacts with the planets for the different configurations (Config.). PCb represents planet b and PCc a possible second planet c.}
	\label{tab6}
	\begin{tabular}{l|cc|c|cc|c} 
		\hline
		Config. & {\bf C1}& \multicolumn{2}{|c|}{\bf C2} & {\bf C3} & \multicolumn{2}{c}{\bf C4}\\
		Impact & PCb	& PCb & PCc & PCb  & PCb & PCc\\
		probability &$\cdot 10^{-4}$&$\cdot 10^{-4}$ &$\cdot 10^{-4}$&$\cdot 10^{-4}$&$\cdot 10^{-4}$&$\cdot 10^{-4}$\\
		\hline
		\multicolumn{7}{l}{Direct impacts}\\
		{\bf R0} &    &    &    &  14  & 12 & 2  \\
		{\bf R1} & 14 & 10 & 4  & 14 & 16 & 2  \\
		{\bf R2} &  2 &  6 &$<$~2& 10 & 2  &$<$~2 \\
		{\bf R3} &  6 &$<$~2&$<$~2& 4 &$<$~2&$<$~2 \\
		\hline
		\multicolumn{7}{l}{Unique MOID impacts}\\
		{\bf R0} &     &     &     & 1832 & 2974 & 2042  \\
		{\bf R1} & 346 & 572 & 390 & 528 & 810 & 460  \\
		{\bf R2} & 172 & 234 & 126 & 514 & 568 & 170 \\
		{\bf R3} & 152 & 178 & 66  & 262 & 234 & 86 \\
		\hline
		
	\end{tabular}
 \end{table}

\begin{table}
    \centering
    \caption{Summary of the water-mass-transport which represents the amount of water transported to the planets given in units of Earth ocean masses for the different configurations (Config.). The water-mass-transport will be influenced by the total mass of the ring (in solar masses), which will depend on the exponent: $\alpha=1.5$
		$M_{tot}$(R0)=4.24$\cdot 10^{-5}$, 
		$M_{tot}$(R1)=5.56$\cdot 10^{-5}$, 
		$M_{tot}$(R2)=3.17$\cdot 10^{-5}$, 
		$M_{tot}$(R3)=2.18$\cdot 10^{-5}$ 
		$M_{\odot}$, and $\alpha=2.2$
		$M_{tot}$(R0)=2.57$\cdot 10^{-5}$, 
		$M_{tot}$(R1)=8.66$\cdot 10^{-6}$, 
		$M_{tot}$(R2)=2.21$\cdot 10^{-6}$, 
		$M_{tot}$(R3)=8.99$\cdot 10^{-7}$}
	\label{tab7}
	\begin{tabular}{l|cc|c|cc|c} 
		\hline
		Config. & {\bf C1}& \multicolumn{2}{|c|}{\bf C2} & {\bf C3} & \multicolumn{2}{c}{\bf C4}\\
		Planet  & PCb	& PCb & PCc & PCb  & PCb & PCc\\
		\hline
		\hline
		\multicolumn{7}{l}{Water transport in Earth oceans by direct impacts}\\
		\multicolumn{7}{l}{with the exponent $\alpha=1.5$}\\
		{\bf R0} &      &       &       & 3.96 & 3.39 & 0.62 \\
		{\bf R1} & 0.52 & 0.37  & 0.15  & 0.52 & 0.59 & 0.07 \\
		{\bf R2} & 0.04 & 0.13  & 0     & 0.21 & 0.04 & 0    \\
		{\bf R3} & 0.09 & 0     & 0     & 0.06 & 0    & 0    \\
		\hline
		\multicolumn{7}{l}{Water transport in Earth oceans by unique MOID impacts}\\
		\multicolumn{7}{l}{with the exponent $\alpha=1.5$}\\
		{\bf R0} &       &        &         & 51.80 & 84.09 & 57.74  \\
		{\bf R1} & 12.81 & 21.18  & 14.44   & 19.56 & 30.00 & 17.04 \\
		{\bf R2} & 3.63  & 4.94   & 2.66    & 10.85 & 11.99 & 3.59  \\
		{\bf R3} & 2.21  & 2.58   & 0.96    & 3.80  & 3.40  & 1.25  \\
		\hline
		\hline
		\multicolumn{7}{l}{Water transport in Earth oceans by direct impacts}\\
		\multicolumn{7}{l}{with the exponent $\alpha=2.2$}\\
		{\bf R0} &      &       &       & 2.40 & 2.05 & 0.38 \\
		{\bf R1} & 0.08 & 0.06  & 0.02  & 0.08 & 0.09 & 0.01 \\
		{\bf R2} & 0.00 & 0.01  & 0     & 0.01 & 0.00 & 0    \\
		{\bf R3} & 0.00 & 0     & 0     & 0.00 & 0    & 0    \\
		\hline
		\multicolumn{7}{l}{Water transport in Earth oceans by unique MOID impacts}\\
		\multicolumn{7}{l}{with the exponent $\alpha=2.2$}\\
		{\bf R0} &      &      &         & 31.36 & 50.90 & 34.95 \\
		{\bf R1} & 2.00 & 3.30 & 2.25    & 3.05  & 4.68  & 2.66  \\
		{\bf R2} & 0.25 & 0.35 & 0.19    & 0.76  & 0.84  & 0.25  \\
		{\bf R3} & 0.09 & 0.11 & 0.04    & 0.16  & 0.14  & 0.05  \\
		\hline
		
	\end{tabular}
 \end{table}


\section{Conclusions} \label{con}

In this paper, we first investigated the dynamical possibility of a second planet in Proxima Centauri.
We excluded an outer gas giant in our studies because we do not expect another planet with Saturn or Jupiter mass as shown in the radial velocity limits of Fig.~\ref{fig0b}.
Therefore, we did several numerical simulations and found that both planets could be stable even for large planetary masses (1, 5, and 10 $M_{\mathrm{Prox}}$) and large eccentricities of the second planet (up to $e_3 = 0.3$).
The 2:1 and 3:1 MMR mark the inner border of the stability maps for all possible eccentricities of PCb as shown in Fig.~\ref{fig1}. For nearby orbits we found the apsidal corotation resonance (ACR) which stabilizes the planetary system for moderate eccentricities of PCc.
From Fig.~\ref{fig1} we can conclude that the second planet can be very close to PCb like in a compact planetary system~\citep{funk10}.

We made several simulations where we changed the distance of the planetesimal cloud to the host star. We also assumed an Oort-cloud like distribution of inclinations for the planetesimal cloud.
In the next step, we investigated the influence of the binary $\alpha$ Centauri on the water transport onto PCb and a possible second planet orbiting in the HZ.
In case that the planet PCb is too dry, maybe a second planet at the outer border of the habitable zone will have a higher probability to be habitable. We found that the water transport onto planets b and c by direct impacts is rare (except region {\bf R0} and {\bf R1}), compared with MOID impacts.
Our simulations showed in addition that planet b would gain more water than planet c, by the help of direct (impacts) or indirect (close encounters) water transport.
The impact probabilities estimated from the direct impacts are summarized in Tab.~\ref{tab6}, and show that the inner region has a larger water transport (especially region {\bf R0}) than the outer ones (see Tab.~\ref{tab7} and Fig.~\ref{fig9}).
We can conclude that $\alpha$ Centauri perturbs the system, but does not increase the influx of comets and therefore the water transport is not significantly larger.
By studying the outer regions we found out that the limit for water transport is around 200~au for Proxima Centauri.
The planetesimal cloud disappeared  almost entirely for region {\bf R5} and completely for region {\bf R6}.
The observations of dust belts around Proxima Centauri strengthen the idea of water transport by icy objects from
debris disks as shown in \cite{Anglada17}. 
We presume that these results could be important for Proxima like systems.

Further investigations have to be done to solve questions like: 
If $\alpha$ Centauri were closer to Proxima Centauri how much would that influence the system.
Could there be an exchange of comets between $\alpha$ Centauri and Proxima Centauri and to which extent would that be?

This work will possibly help to understand the cosmogony of Proxima Centauri and the dynamics of water transport in M-star systems in general; however, further investigations will be necessary.


\section*{Acknowledgements}

The authors thank Prof. Giovanni Valsecchi and Prof. Christos Efthymiopoulos for their valuable comments.
R. Schwarz wants to acknowledge the support by the Austrian FWF projects P23810-N16, S11608-N16 and S11603-N16.
E. Pilat-Lohinger, D. Bancelin, and {\'A}. Bazs{\'o} want to acknowledge the support from the Austrian FWF project S11608-N16.
R. Dvorak, B. Loibnegger, and T. Maindl want to acknowledge the support from the Austrian FWF project S11603-N16.
K. Kislyakova wants to achknowledge the support from the FWF project S11607-N16.
We thank the High Performance Computing Resources team at New York University Abu Dhabi and especially Jorge Naranjo for helping us carry out our simulations.






\bibliographystyle{mnras}
\bibliography{lit}



\appendix
\section{stability}

To confirm our results in section~\ref{res1} (star-planet-planet) we used the stability catalogue 
``ExoCat''~\citep[after][]{sandor07}. 
The application ``ExoCat'' allows to check whether a newly discovered planet in a single-star single-planet system moves inside a stable region, using results of the work of~\cite{sandor07} and \cite{elke11}. The dynamical model
was the restricted three-body problem, which is only an approximation to the full three body model used for our stability maps. 
The simulations from ``ExoCat'' used the relative Lyapunov indicator (RLI) to investigate the stability of the system.
The eccentricity of the planet PCb was set to $\mathrm{e_2}$=0 for all stability maps.
In that way we looked for the stability of an outer planet (PCc) as shown in Fig.~\ref{fig1}.
We compared the stability plots for two different bracketing mass ratios from the catalogue: $\mu = 1 \times 10^{-4}$ and $2 \times 10^{-4}$ ($\mu = m_{2} / (m_{1} + m_{2}$), and our calculations for 5~$M_{PCb}$ corresponding to
$\mu=1.7 \times 10^{-4}$. The semi-major axis in the figure is normalized to unit distance for PCb, i.e. $a_3=a/a_2$.

\begin{figure*}
    \centering
    \includegraphics[width=10.5cm]{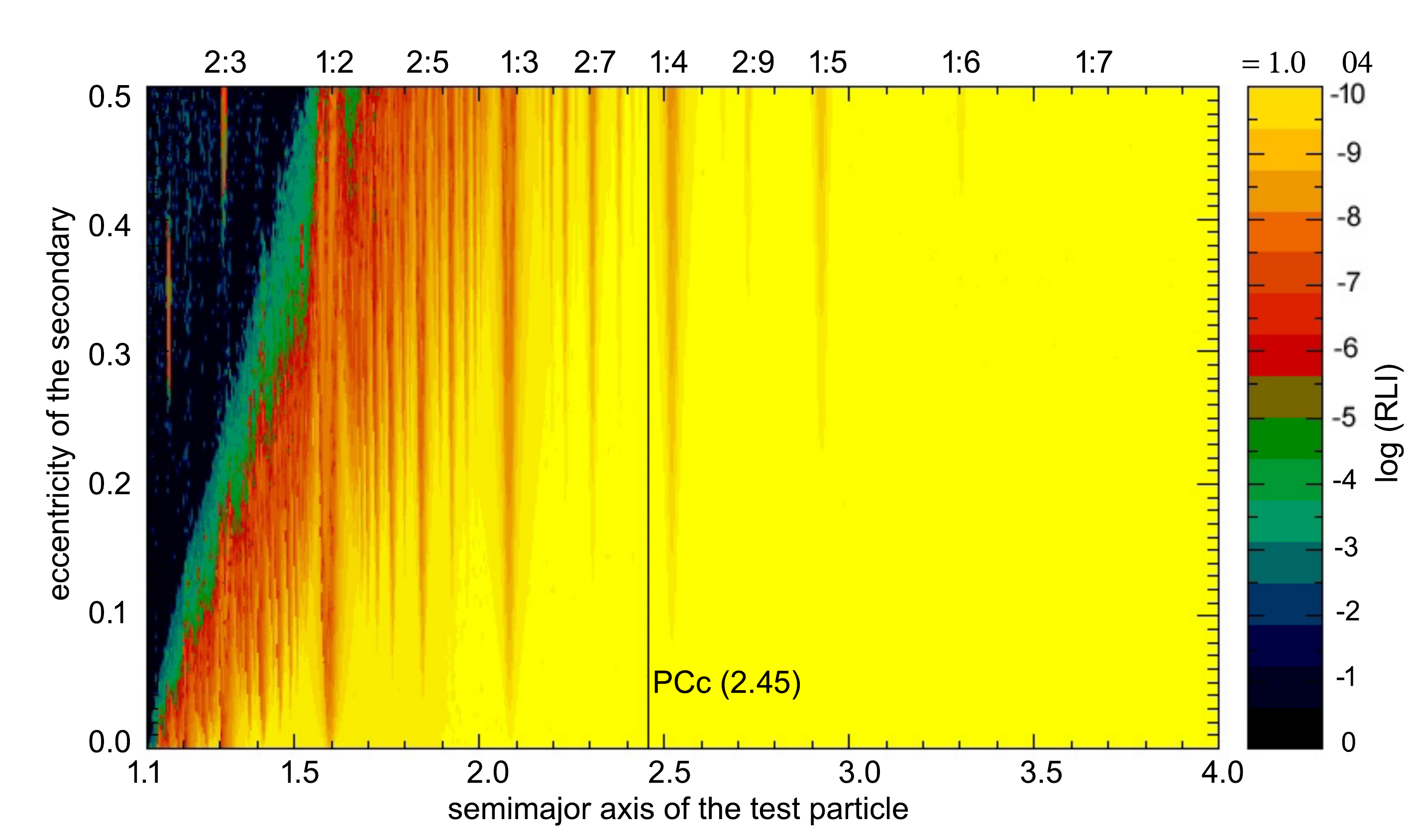}\\
    \includegraphics[width=10.5cm]{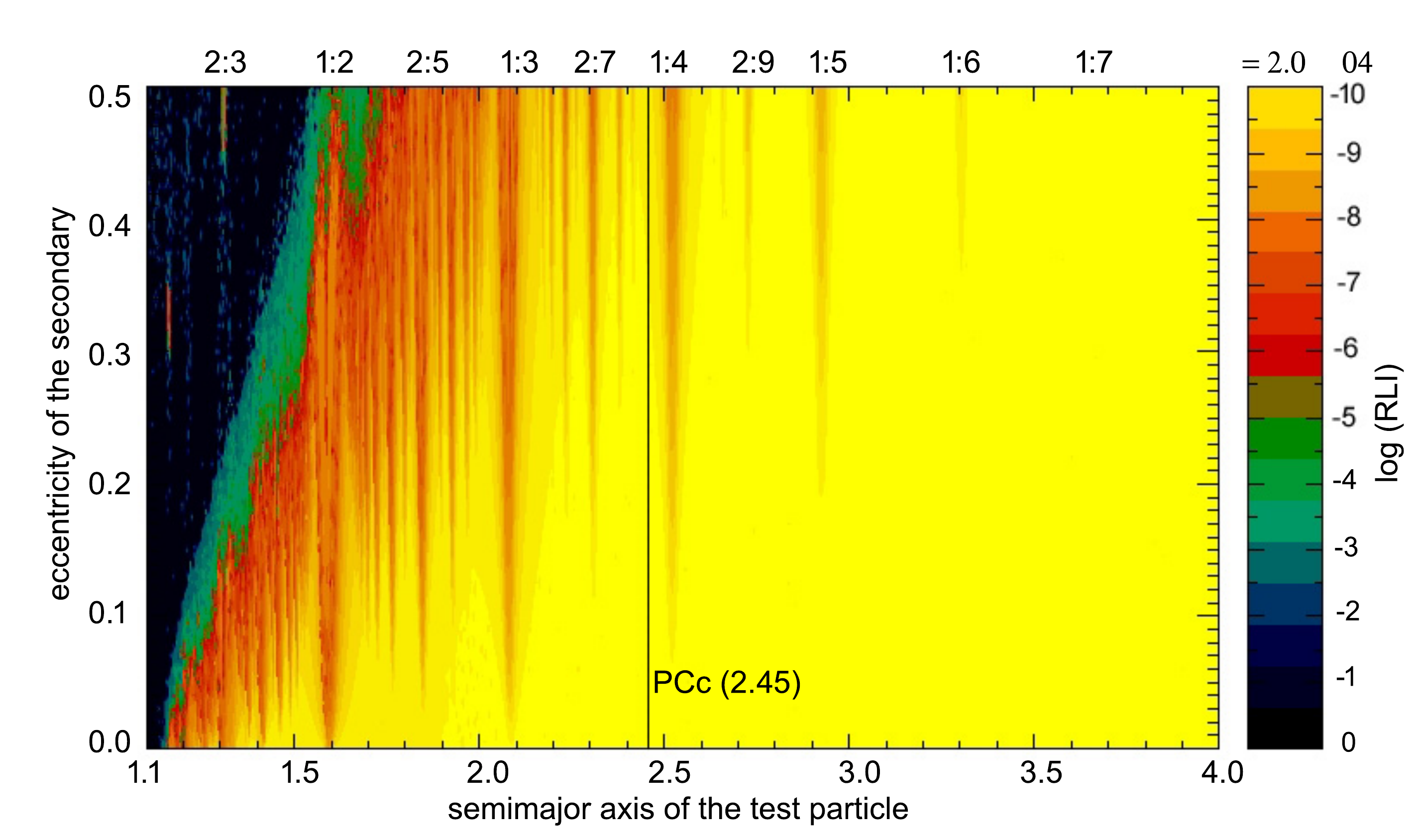}\\
    \includegraphics[width=11.0cm]{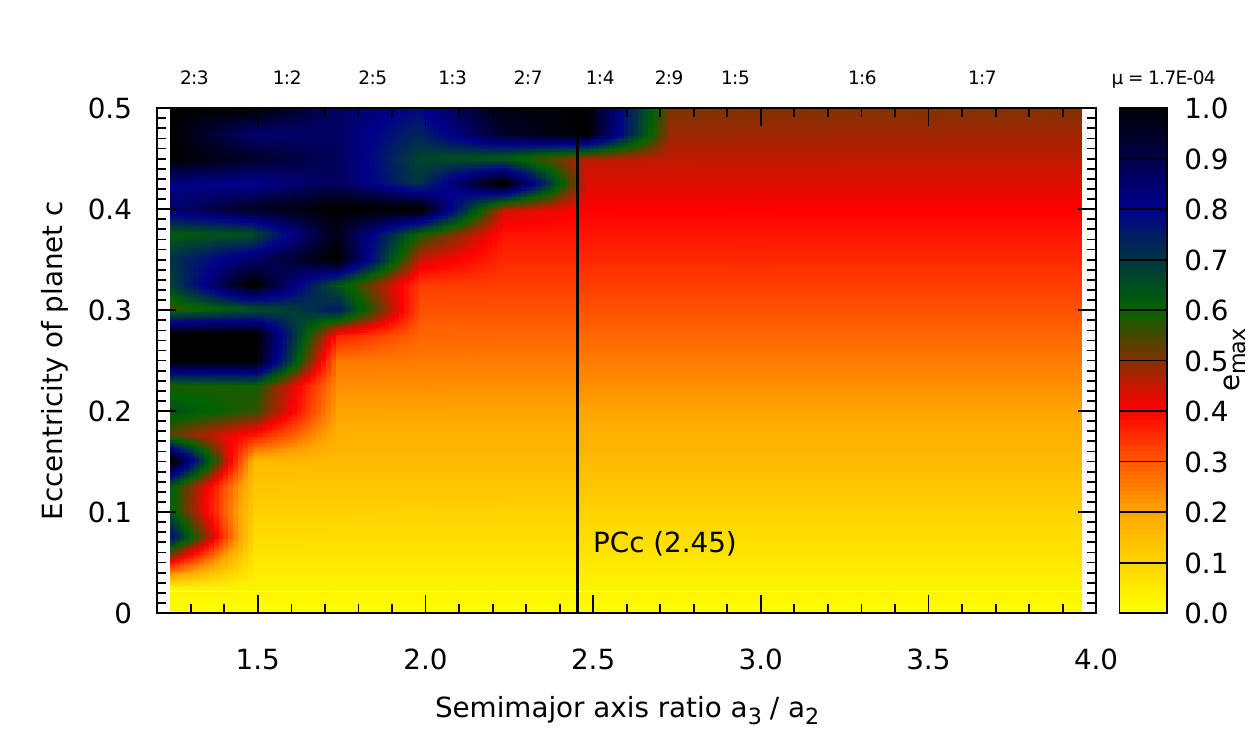}\\
    \caption{
The upper and middle graphs show the stability maps according to "ExoCat" for an additional (massless) planet PCc.
These maps display the size of the stable region (negative numbers are more regular, values around zero are chaotic) for two different mass ratios ($\mu=10^{-4}$ and $\mu=2 \cdot 10^{-4}$) of PCb relative to the host star.
The lower graph presents the stability map of PCc from Fig.~\ref{fig1} for equal masses of the planets: 
$m_2$=$m_3$=5$\cdot M_\mathrm{Prox}$. The colour code presents the values of $e_\mathrm{max}$ (lower graph) and the chaos indicator RLI (upper graphs) as given in the figures.}	
    \label{fig4}
\end{figure*} 

\begin{figure*}
    \centering
    \includegraphics[width=10.5cm]{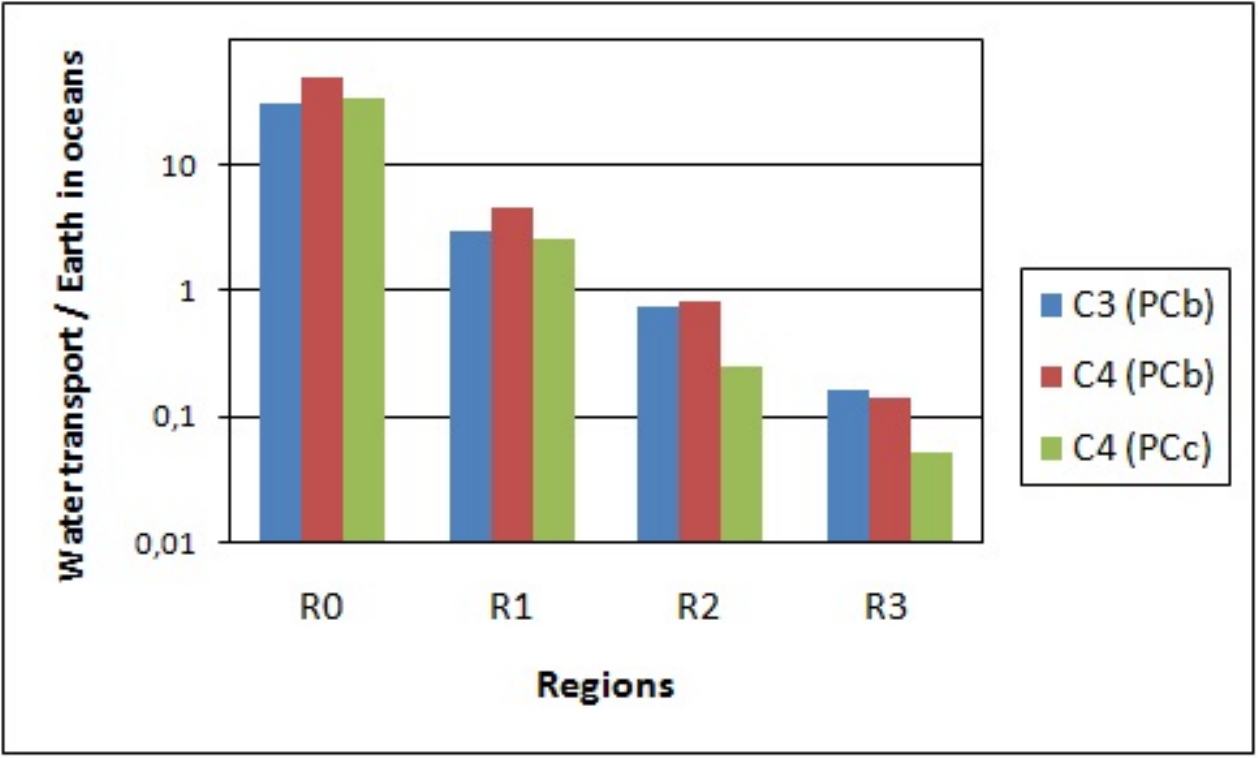}\\
    \caption{
The figure is a conclusion of Tab.\ref{tab7} and shows that the water transport is most effective for the region 
R0 (1-4~au) close to the planet(s).}	
    \label{fig9}
\end{figure*} 

\begin{figure*}
    \centering
    \includegraphics[width=10.5cm]{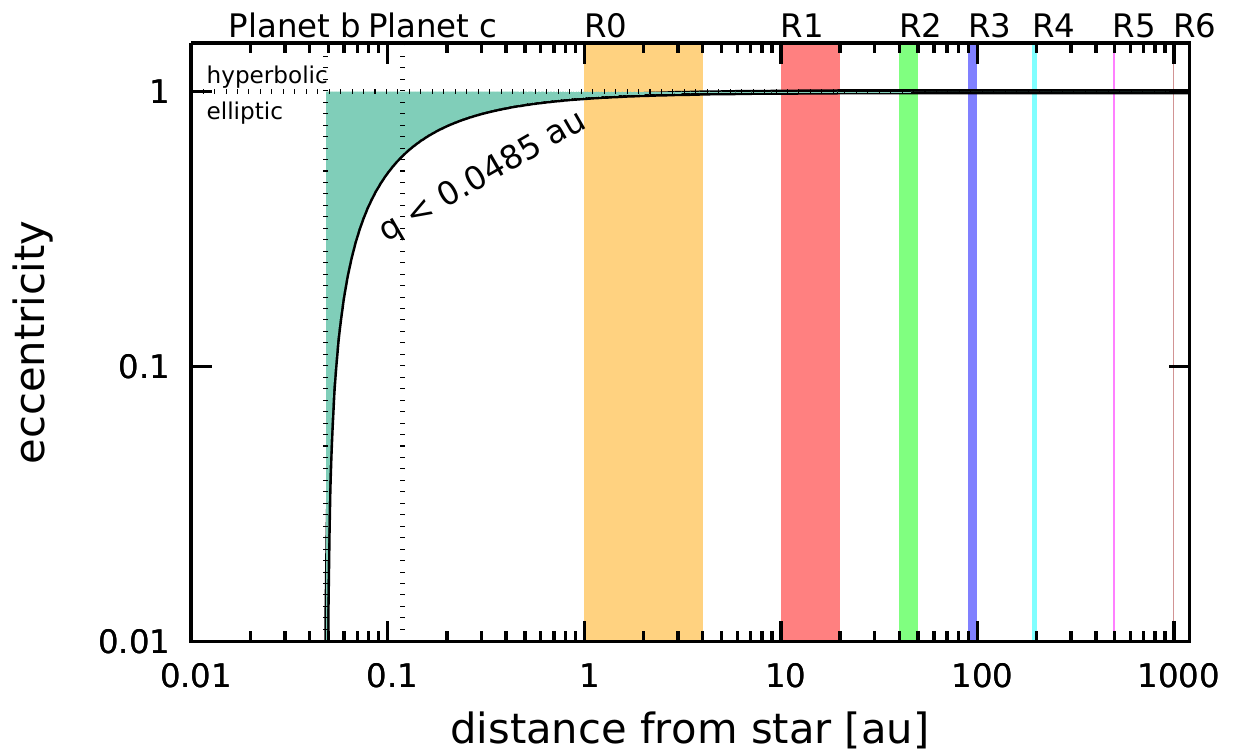}\\
    \caption{
The figure show the range of the planetesimal clouds (different colours) and the perihelion distance of the comets for the different regions. We set the initial eccentricity of the comets high enough such that the perihelion distance q $<$ 0.0485~au in any case (typically e $>$ 0.95).
}	
    \label{fig10}
\end{figure*}



\bsp 
\label{lastpage}
\end{document}